\documentclass[a4paper,11pt,fleqn]{article}

\usepackage[centertags]{amsmath}
\usepackage{amssymb,bm}
\usepackage{amsfonts}
\usepackage{csvsimple}

\usepackage{booktabs} 
\usepackage{pgfplotstable} 

\usepackage{graphicx}
\usepackage{color}
\usepackage{placeins}
\usepackage[square]{natbib}

\usepackage{multirow}
\usepackage{booktabs}

\usepackage{array,multirow,makecell}
\newcolumntype{R}[1]{>{\raggedleft\arraybackslash }b{#1}}
\newcolumntype{L}[1]{>{\raggedright\arraybackslash }b{#1}}
\newcolumntype{C}[1]{>{\centering\arraybackslash }b{#1}}

\pgfplotsset{compat=1.17}

\newcommand{\dt}{\delta t}

\newcommand{\DT}{\Delta T}

\newcommand{\cdf}{\operatorname{cdf}}

\newcommand{\VaR}{\operatorname{VaR}}
\newcommand{\ES} {\operatorname{ES}}

\begin{document}


\begin{center}
	{ \bf \huge Multivariate backtests and copulas \\[1ex] for risk evaluation}

	\vspace{5ex}
	{\bf\large
		Boris David and Gilles Zumbach
	}\\[2ex]
	\parbox{0.4\textwidth}{\renewcommand{\baselinestretch}{1.0}\normalsize
		Edgelab\\
		Avenue de la Rasude 5\\
		1006 Lausanne\\
		Switzerland
	}
	\\[3ex]
	\parbox{0.4\textwidth}{\renewcommand{\baselinestretch}{1.0}\normalsize
		bdavid@edgelab.ch\\
		gzumbach@edgelab.ch 

		\vspace*{4ex}
		October 17, 2023  
	}


\begin{abstract}
Risk evaluation is a forecast, and its validity must be backtested. Probability distribution forecasts are used in this work and allow for more powerful validations compared to point forecasts. Our aim is to use bivariate copulas in order to characterize the in-sample copulas and to validate out-of-sample a bivariate forecast. For both set-ups, probability integral transforms (PIT) and Rosenblatt transforms are used to map the problem into an independent copula. For this simple copula, statistical tests can be applied to validate the choice of the in-sample copula or the validity of the bivariate forecast. The salient results are that a Student copula describes well the dependencies between financial time series (regardless of the correlation), and that the bivariate forecasts provided by a risk methodology based on historical innovations performs correctly out-of-sample. A prerequisite is to remove the heteroskedasticity in order to have stationary time series, in this work a long-memory ARCH volatility model is used.

\end{abstract}
\end{center}
\vspace{2ex}


\newpage


\section{Introduction}

An important research activity in finance concerns the construction of models for financial time series, both univariate and multivariate.
Having realistic models is very important for many practical applications in the industry, and this paper focuses on multivariate risk evaluations.
Crucially, risk estimation is a forecast, namely to evaluate now, say at time $t$, the magnitudes for the possible losses for an asset or a portfolio between $t$ and sometime later, say at time $t+\DT$ where $\DT$ is a desired risk horizon.

Investing involves a large dimensional space of assets, with a diverse portfolio made of positions on those assets.
Computing risk for a portfolio adds the complexity of the multivariate dependencies between assets.
Most assets are positively correlated, the typical example being stocks and stock indexes.
We want to characterize the dependencies beyond linear correlations, and the information about the full dependencies are contained in copulas.
Getting the asset risks as well as the dependencies correctly implies that the composite risk of a portfolio is also correct.

Beyond the aggregated risk of his portfolio, a portfolio manager does not only want to get its risk measures, he/she needs to act in order to modify its risk toward a desired objective.
Such decisions are taken by using finer diagnostics, in particular risk contributions, namely the contributions of each position to the total risk of the portfolio.
Computing correctly the risk contributions is fundamentally a multivariate problem, involving each position considered in the portfolio.
Another method frequently used in the industry is to decompose the risk along a typology, for example currencies or asset types.
For such a decomposition to be correct, the dependencies between assets should be fully identified so that the aggregation into buckets gives a quantitatively correct decomposition.
These examples show that risk evaluation for a portfolio is fundamentally multivariate, and the co-dependencies are important both for risk aggregation and for the disaggregation involved in the evaluation of finer diagnostics.

Financial risks are quantified usually using tail risk measures, like value-at-risk ($\VaR$) or expected shortfall ($\ES$) at a given confidence level $\alpha$.
This reduces the risk evaluation to a point forecast, producing a single number at each date $t$.
Yet, it is much more powerful for a risk evaluation to produce a density forecast, namely to compute the probability density function $p(l)$ for the losses $l$.
Much more information is contained in $p(l)$, in particular all the risk measures can be computed at any desired confidence level $\alpha$.
Furthermore, the validation of a risk evaluation in a backtest allows for better statistical tests, essentially comparing the sequence of time-dependent forecasts for $p_t(l)$ with the realized loss $l(t+\DT)$ between $t$ and $t+\DT$ by using a probability integral transform (PIT).

Remains the problem of building the forecast $p(l)$.
A process for the time evolution of the assets is a good foundation to build such forecasts.
Any financial forecast should be rooted in stylized facts, namely statistical regularities observed in most time series, and in dynamic models for the time evolution of the prices, namely in processes.
The base model is a random walk for the prices, but clear deviations are observed empirically from the simplest constant volatility process.
Quantitatively, the most important stylized fact is the clustering of volatility, also called heteroskedasticity.
This clustering is precisely what allows for a forecast, since period of low (resp. high) volatility are likely to be followed by low (resp. high) volatility.
A quadratic long-memory ARCH (LM-ARCH) process is used in this paper to model the dynamics of the volatility, leading to simple and ``universal'' volatility forecasts.

An important pre-requisite for any statistical test on time series is that the data must be stationary.
Because of the volatility dynamics, returns are not stationary.
Some care needs to be taken in order to discount for the volatility dynamics and to obtain stationary random variables.
The returns discounted for the volatility dynamics are called innovations (since the volatility is predictable to a good extent, the returns after removing the predictable part are fundamentally the new information in the process).
The historical innovations are serially independent, and follow a fixed time-independent distribution (at least to a good approximation).
Both stylized facts are the main ingredients to build a risk evaluation: the volatility forecast fixes the size of the distribution $p(l)$, and the shape of $p(l)$ is given by the stationary distribution of the innovations.
These forecasts are based on the available information up to a given time $t$, say for example, to measure the current level of volatility.
If an analytical form is needed for the univariate probability distributions, a Student distribution provides for a simple and accurate shape for most assets.

The present contribution is based on a risk algorithm used at an industrial scale and for large multivariate portfolios.
In particular, the volatility clustering is captured by a LM-ARCH process and the distributions of the innovations is sampled from historical time series (i.e. an analytical distribution is not used).
Let us emphasize that our goal is not to build a new risk methodology using copulas, but to study an existing algorithm in order to check that it performs as desired.
Beyond this particular risk algorithm, the goal is to construct and evaluate diagnostics for multivariate risk computations, and in particular to bi-variate copulas.

Since risk is a forecast, it is very important to validate a risk algorithm using backtests.
The key idea of backtest is to replay past data, computing risk forecasts using the information up to a given time $t$, then to peek at the later data to check if the successive forecasts are statistically correct.
An important literature exists for the univariate testing of risk forecast, but is lacking for the multivariate backtesting.

This paper focuses on characterizing the bivariate copula between time series for the innovations.
In a backtest evaluation, the in-sample and out-of-sample cases must be carefully separated and distinguished.
First, the in-sample characterization of the copula is studied.
This is the usual application of copulas, aiming at assessing the long-term dependencies in order to build multivariate processes.
As shown by a statistical test,  Student copulas provide for an accurate description of the dependencies, but with the degrees of freedom depending weakly on the asset pair.
Second, an out-of-sample set-up is used to validate a multivariate forecast.
In this case, the sequence of forecasts is in the form of a multivariate probability distribution, and the realized vectors of losses should be drawn according to the distributions.
Using a copula allows to check the validity of such multivariate forecasts.
A risk methodology based on a trailing sample of historical innovations provides for good bivariate forecasts.
Because the tests involve conditional probability, and because of the sample size used to construct the forecast, it is not possible to test for copula with dimensions higher than 2.
Because of the limitation due to the time series length, only a risk horizon of 1 day can be tested.

Many topics have been mentioned in the above road map, with a large literature for all of them.
A general reference with a mathematical approach on risk evaluation is \citep{Embrechts.2015}, which covers many subjects such as tail risk measures, ARCH processes and copulas.
Another general reference is \citep{JondeauPoonRockinger.2007}, covering similar subjects but with an econometric approach.
Both references provide for excellent introductions to these topics, with ample references to the primary literature.

The base model for the price process is the normal random walk introduced by \citep{Bachelier.1900}.
The clustering of the volatility was recognized in the '80s, in particular with the introduction of the GARCH processes by \citep{Engle.1982} and \citep{EngleBollerslev.1986, Bollerslev.1986}.
Many processes have been introduced along these lines, with the goal of providing for an accurate description of the volatility dynamics, see for example the reviews \citep{AndersenBollerslevChristoffersenDiebold.2006, HansenLunde.2005, Poon.2005}.
For risk evaluation, the goal is to have a simple and robust volatility formula providing for good enough estimates for a very large panel of securities and indexes.
It is an engineering task, where the major stylized properties should be included but some details have to be left out for simplicity.
Our approach is to use a LM-ARCH volatility forecast, based on the process introduced in \citep{Zumbach.LongMemory} (see also \citep{Corsi.2009} for a related approach).
The few parameters in the process have been studied then (see also \citep{Zumbach.RM2006_fullReport}), and do not need to be estimated.
This is an important advantage for large scale applications.

Regarding multivariate ARCH processes, our study is in the line of a CCC-ARCH (Constant Conditional Correlation) model \citep{Bollerslev.1990}, but instead of the correlations for the returns, it is the dependencies for the innovations that are studied.
As the volatility is taken care on a time series basis, the correlation values are not needed as would be for a process, but are evaluated with the historical time series of the innovations (i.e. the return discounted by the volatility).
The same approach is used for the dependencies using copulas.

A thread in the literature is studying the DCC-ARCH (Dynamic Conditional Correlation) process for the returns.
Let us emphasize that our base variables are the innovations, for which more stable dependency structures can be expected.
In this direction, the in-sample evaluations implicitly assume a constant dependency structure, while the out-of-sample computations on a trailing sample include naturally a time dependency in the copulas.
The consistency of the results between both settings points to a small time dependency in the copula between the innovations, namely the dependency structures appear stable with time.
Yet, such difficult questions would deserve a specific study.

The backtest of risk evaluation is another topic with a large literature.
Most papers study VaR exceedances at a given probability $\alpha$, in relation with a point forecast.
The obvious test is to count the number of exceedances above VaR, and the time independence of the exceedances is another important statistical test, see for example \citep{Kupiec.1995, CrnkovicDrachman.1996, Barbachan.2006, ChristoffersenPelletier.2004, Kratz.2016} on these topics, or \citep{Christoffersen.2010, Roccioletti.2016} for recent reviews.
Regarding point forecasts, some recent extensions concern the multivariate aspect of the exceedances, see \citep{Danciulescu.2010, WiedWeissZiggel.2015}.
Tests based on a probability distribution forecast and a probability integral transform (PIT) have a long history in statistics, and in other fields like weather forecast \citep{Hamill.2001, Diks.2008}.
PIT have been introduced in finance by \citep{DieboldGuntherTay.1998, Zumbach.backtesting}, and used extensively in \citep{Zumbach.TileTest}.
The present work uses a probability forecast, with a subsequent PIT transform in the univariate case and its extension in the multivariate case known as the Rosenblatt transform \citep{Rosenblatt.1952}.

Correlations, measures of dependence and copulas are yet other topics with a large literature.
The classical measures of dependence (Pearson's correlation, Kendall's tau, Spearman's rho) are reviewed in many books and references, see for example \citep{Spearman.1987, GibbonsChakraborti.2014, SchoberBoerSchwarte.2018} for some recent publications.
Regarding copulas, see for example \citep{BouyeDurrlemanNikeghbaliRibouletRoncalli.2000, Nelsen.2010, Embrechts.2015} for general introductory references, or \citep{HofertKojadinovicMaechlerYan.2018} for an in-depth reference with many examples using the language R.
A use of copulas to study dependencies in finance can be found in \citep{Kresta.2012}, \citep{HofertMachler.2014} and \citep{ChicheporticheBouchaud.2015}, but these studies have been done using returns which are not stationary.
Noticeably, \citep{HofertMachler.2014} are using an (in-sample) pair-wise approach to high-dimensional problems, conceptually similar to this paper.
\citep{ChicheporticheBouchaud.2015} studied the return copula for a set of stocks, pointing to a complex structure, and proposing a factor model for the volatilities in order to explain the dependencies.
\citep{FritzschTimphusWeiss.2021}  uses a large panel of risk forecasts, using many variants of GARCH(1,1) to discount for the (univariate) volatility dynamics and different models for the copula of the innovations, with the aim to quantify the model risk.
One of their conclusions is that the copula is an important component of a risk model, while the model for the marginals has less impact.

Testing for copulas, more precisely measuring the relation between a theoretical copula and an empirical sample, is another field with an existing literature.
\citep{GenestRemillard.2004, GenestFavreBeliveauJacques.2007} have presented a test measuring an $L^2$ distance between two copulas.
After a Rosenblatt transform, this GR-test can be used to quantify if an empirical copula is uniform.
Unfortunately, in an out-of-sample setting, this test depends on the original distribution.
In particular, the $p$-value function is depending on the correlation, making this test unsuitable for out-of-sample validation.
Another test is used both in- and out-of-sample, based on an idea from \citep{Knuth.1998} and applied in finance in \citep{Zumbach.TileTest}.
This tile-test is based on a regular tilling on the unit square to measure deviation from uniformity, and testing for the standard deviation from the average number of points in each tile.

To our knowledge, the main contributions of this paper are the following.
As mentioned, most published statistical analysis or backtests are based on the (non-stationary) returns.
Innovations are used instead as the base variable for the tests, and the in-sample analysis of multivariate stationary innovations is new.
Then, the out-of-sample backtest of multivariate risk evaluations using copula is new, and this is the main objective of this work.
This analysis is based on a Rosenblatt transform computed on sampled distributions, which is also new.
Two statistical tests for uniformity are used, the GR-test and tile-test, but only the tile-test has good properties with respect to out-of-sample correlated copulas.
The tile-test, and the application on an out-of-sample setting is new.
Then, in a large cross-sectional study, in-sample and out-sample copulas are found to have similar properties, with an inverse dependency between correlation and tail index (i.e. pairs with larger correlations have Student copulas with smaller degrees of freedom).
Finally, beyond multivariate normal, a simple multivariate Student distribution for the innovations is suggested.

The present computations have been made using the language R, in particular the powerful package ``copula'', see \citep{HofertKojadinovicMaechlerYan.2016,HofertKojadinovicMaechlerYan.2018}.
We have used only bivariate copulas, first because visual control can be made, and second because the out-of sample setting does not allow to compute conditional probability with enough points when the copula dimension is larger than 2.
In order to check that a model's copula is indeed describing correctly some empirical data, a Rosenblatt transform is used to map the data so that the copula becomes an independent copula if the model is correct.
Among so many functions, the R package implements the GR-test \citep{GenestRemillard.2004, GenestFavreBeliveauJacques.2007} allowing to test that a copula is an independent copula.
This is used in the in-sample section, but the $p$-values are correlation-dependent in the out-of-sample setting, making it unsuitable in this context.
Instead, the tile-test is used both in- and out-of-sample.
For this test, the $p$-values are computed with Monte Carlo simulations.

In order to illustrate the different computations for the in-sample model and out-of-sample backtest, the copula between  the DAX and Dow Jones Industrial Average (DJIA) will be used.
For the corresponding figures, the data span returns and innovations from 2006-01-01 to 2021-05-13, for a data length of 3791 points.
This sample length is large enough to give a clear view of a sampled copula, but not having to many points to cover densely a unit square.
For the cross-sectional empirical evaluations, one wider data set has been used, composed of various indexes and FX, with longer samples for the empirical tests.

The presentation proceeds as follows.
Section~\ref{sec:stationarity} explains the notion of stationarity, its importance in the context of finance and the discounting of the volatility to obtain stationary time series.
Sec.~\ref{sec:inSample} introduces the in-sample set-up and some key concepts such as copulas, Rosenblatt transform, uniform copulas and the tests for independence.
These computations are applied in Sec.~\ref{sec:inSampleResults} on the cross-sectional data set in order to characterize the long-term dependencies between time series.
The out-of-sample forecast set-up is presented in Sec.~\ref{sec:out-of_sample}, extending the in-sample concepts from the section~\ref{sec:inSample} and with the required modifications for the $p$-values.
Empirical results about risk forecast are presented in Sec.~\ref{sec:out-of-sampleResults}, before the conclusions.

\section{Stationarity}
\label{sec:stationarity}

In time series analysis, an essential mathematical prerequisite is stationarity.
A time series is said to be stationary if the distributions from which the observations are sampled at each date $t$ are identical.
This implies in particular that a time series has constant mean and variance over time.
More detailed definitions about stationarity and its importance in forecasting has been discussed in \citep{Greunen.stationarity}.

Thanks to the seminal works on ARCH processes, the time series of returns is known to have a time dependency in its variance, called heteroskedasticity.
This corresponds to periods where the corresponding financial markets are quiet, volatile, or in a crisis.
The time dependency of the volatility is large, with possibly a factor 10 between quiet and crisis periods, and this can be visualized simply by plotting the daily returns from any time series over a long enough span, say 20 years.
Such a large effect cannot be ignored, and a prerequisite for a sound statistical analysis is to discount for the time dependent volatility in order to have time series that are homoskedastic.

We are using the ARCH framework from Engle and Bollerslev, using the long memory version of the process presented in \citep{Zumbach.LongMemory}, called LM-ARCH.
Daily data are used in this study, with $\dt$ the one-day time step.
In order to model the dynamics of the volatility in a process, the returns at time $t+\dt$ are expressed as
\begin{equation}
	r(t+\dt) = \sigma(t)\,\varepsilon(t+\dt)
	\label{eq:ARCH}
\end{equation}
where $\sigma(t)$ corresponds to the volatility, evaluated using the past returns up to time $t$.
It is a forecast, made at time $t$, for the volatility in the period $t$ to $t+\dt$.
The innovation $\varepsilon(t+\dt)$ is assumed to be iid, with a stationary distribution $p(\varepsilon)$ with a unit variance.

A drift forecast $\mu(t)$ can be added in the relation \eqref{eq:ARCH}, but it is small, difficult to estimate and sample dependent.
Since financial time series are dominated by the volatility at short time scale, the drift term can be neglected.
The relative importance of drift versus volatility, as $\DT$ is increased, is rooted in their respective scaling, as $\DT$ for the drift and as $\sqrt{\DT}$ for the volatility.
For most time series, the cross-over from volatility dominated at short time, to drift dominated at long time occurs at a time scale of a few years.
This order of magnitude for the cross-over time indicates that the drifts are negligible compared to the volatility at the scale of 1 day.
When neglecting the drift term for the returns, the distribution $p(\varepsilon)$ can have a mean different from zero, but much smaller than 1 in practice.

Plenty of models are available to compute $\sigma(t)$ from the past returns, say for example the GARCH(1,1) process.
The LM-ARCH process is a simple quadratic form of the past returns, with weights decaying slowly with increasing lags from $t$, capturing the gradual but slow decay of the information as the time passes by.
It depends on very few parameters that can be taken as fixed for all time series.
Essentially, the LM-ARCH volatility estimate is a very good, simple and universal workhorse to describe the heteroskedasticity, and does not need to be estimated on each time series.
Its efficiency to remove volatility clustering has been investigated in depth in \citep{Zumbach.TileTest}.

Eq.~\eqref{eq:ARCH} is written as a process, with $\varepsilon(t)$ drawn from $p(\varepsilon)$ at each time.
With historical time series for the returns, equation \eqref{eq:ARCH} can be expressed for the innovations
\begin{equation}
	\varepsilon(t+\dt) = \frac{r(t+\dt)}{\sigma(t)}
	\label{eq:innovation}
\end{equation}
where the right-hand side is a function of the past returns up to $t+\dt$  for the return and up to $t$ for $\sigma$.
This equation allows to compute the time series of realized innovations (or historical innovations).
If the one-step volatility forecast is good, the heteroskedasticity is expected to be gone, and the distribution for the innovations to be stationary.

It can be checked that this is indeed the case to a very good approximation, using a stringent statistical test (see \cite{Zumbach.TileTest} and references therein).
For this paper, we take for granted that the distribution of $\varepsilon$ is iid.
Notice that no particular shapes for the distributions are assumed, but in practice a Student distribution gives a good description for the realized innovations for many assets and indexes.

\begin{figure}[ht]
	\centering
	\begin{minipage}{.45\textwidth}
		\centering
		\includegraphics[width=1\linewidth]{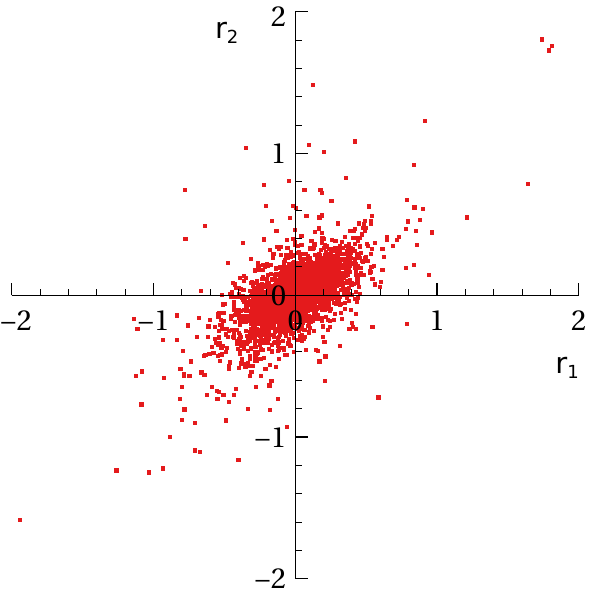} \\
	\end{minipage}%
	\hspace{1cm}
	\begin{minipage}{.45\textwidth}
		\centering
		\includegraphics[width=1\linewidth]{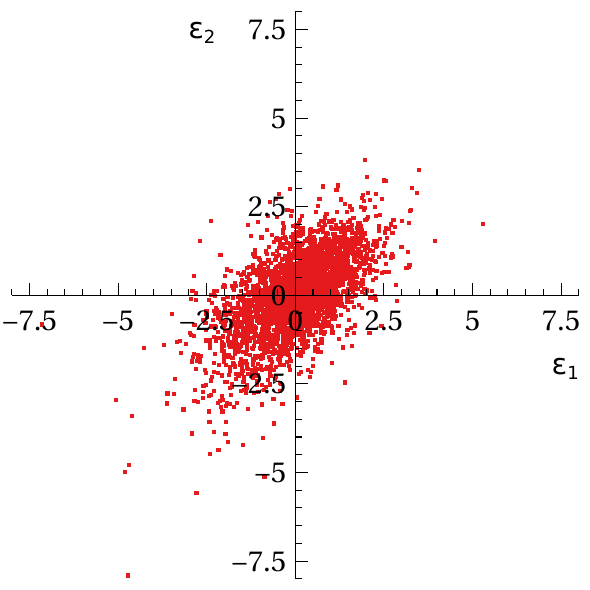}\\
	\end{minipage}
	\caption{The returns (left-side) and the innovations (right-side) of DAX (x-axis) and DJIA (y-axis) plotted pairwise.
		The innovations are computed according to formula~\eqref{eq:innovation}.}
	\label{fig:datasets}
\end{figure}

For our selected example of DAX versus DJIA, the bivariate distribution for the returns and the innovations are plotted in Fig.~\ref{fig:datasets}.
Our focus for this paper is on the multivariate aspects of the dependencies in finance, both in-sample and out-of-sample in a risk forecast setting.
As the innovations are (assumed to be) stationary, this will be our base variables for this study.
For the same reason, the risk forecast presented in Sec.~\ref{sec:out-of_sample} is constructed on the stationarity of the innovations.

\section{In-sample evaluation of copula}
\label{sec:inSample}

\subsection{The in-sample set-up}

The common set-up for copulas is to study the in-sample dependencies between variables, computing the cumulative distributions and combining the pseudo-observations obtained in the sample from the innovations.
``In-sample'' means that a model is estimated and validated on the full data set, and the goal is to characterize the multivariate dependencies as they occurred in the chosen sample.
Our further goal is to validate distributional forecasts, that are time-dependent.
This requires a slightly different set-up, called ``out-of-sample'', in order to backtest the forecast.
The presentation in this section is done in a way to extend simply out-of-sample in Sec.~\ref{sec:out-of_sample}.

The mathematical set-up assumes stationary random variables.
As discussed in the previous section, the heteroskedasticity of the returns invalidates any stationarity assumptions.
Instead, the innovations $\varepsilon$ are used in this contribution, with essentially no lagged dependencies and stationary distributions (see e.g. \cite{Zumbach.TileTest}).

The appropriate mathematical objects for studying dependencies are copulas.
We recapitulate here the usual set-up in order to fix some notations, empirical definitions and statistical tests, and to make clear the differences with the out-of-sample copula used in sections~\ref{sec:out-of_sample} and \ref{sec:out-of-sampleResults}.
A copula is a $d$-dimensional multivariate distribution function on $[0,1]^d$ with uniform marginals
\begin{equation*}
	C: [0,1]^d \rightarrow [0,1]
\end{equation*}
They provide for a universal representation of the whole dependencies between $d$ random variables.

The in-sample computations can be summarized as follows.
\begin{itemize}\vspace{-1.5ex}
	\item For each considered asset $i$, compute the marginal empirical cumulative distribution function (cdf) $P_i(\varepsilon)$ for the innovations using the full sample of data.

	\item Use this empirical cdf to perform a probability integral transform (PIT) of the innovations
	\begin{equation}
		z_i(t) = P_i(\varepsilon_i(t)).
	\end{equation}
	The variable $z$ is the cumulative probability of the quantile $\varepsilon$.

	In the in-sample setting, this procedure is equivalent to using the scaled ranks of the innovations and is the usual construction considered in the copula literature.
	In the out-of-sample setting, the empirical cumulative distribution is replaced by a forecast $\widetilde{P}_t$ (see Eq.~\eqref{eq:probtileOutOfSample}), and $z$ is not equivalent to the scaled rank due to the time dependency observed in the forecast for the distributions.
	For these reasons, we named $z$ the \textit{probtiles} in order to make clear the distinction with the ``pseudo-observations'' defined in-sample.

	In-sample, the probtiles are the ``pseudo-observations'' and are uniformly distributed random variables over $[0,1]$ by construction.
	Out-of-sample, the distribution for $z$ is uniform only if the forecast $\widetilde{P}_t$ is correct for $\varepsilon(t+\DT)$, see Sec.~\ref{sec:out-of_sample}.

	\item The $d$-dimensional vectors $(z_1(t),...,z_d(t))$ are a sampling from the copula on this data set.
	We call it the \textit{in-copula}, in order to differentiate from the \textit{out-copula} used in a forecast context in sections \ref{sec:out-of_sample} and \ref{sec:out-of-sampleResults}.
\end{itemize}

\begin{figure}[ht]
	\centering
	\hspace*{-0.4cm}
	\begin{minipage}{.5\textwidth}
		\centering
		\includegraphics[width=1\linewidth]{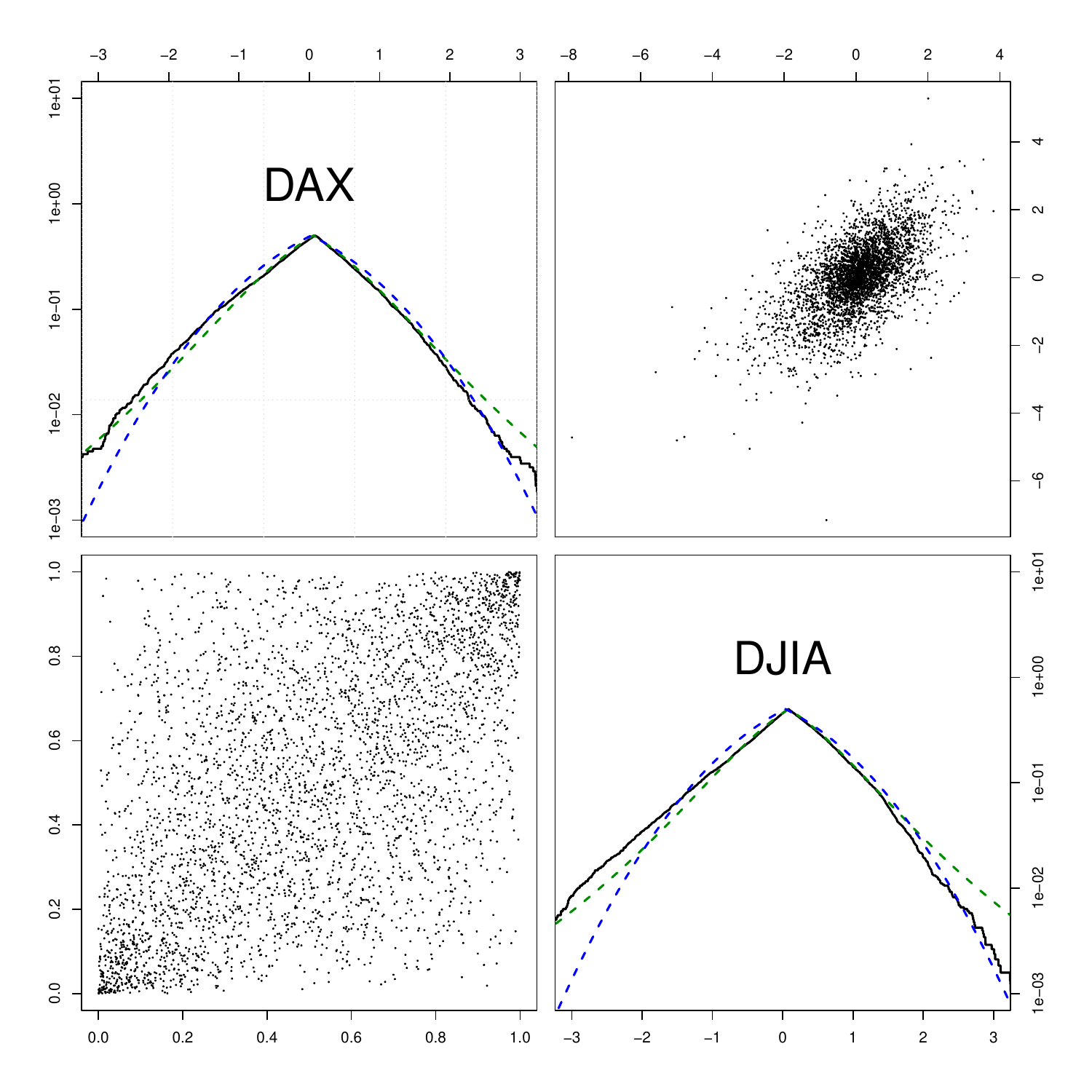} \\
		\label{fig:test1}
	\end{minipage}%
	\hspace*{0.5cm}
	\begin{minipage}{.45\textwidth}
		\centering
		\includegraphics[width=1\linewidth]{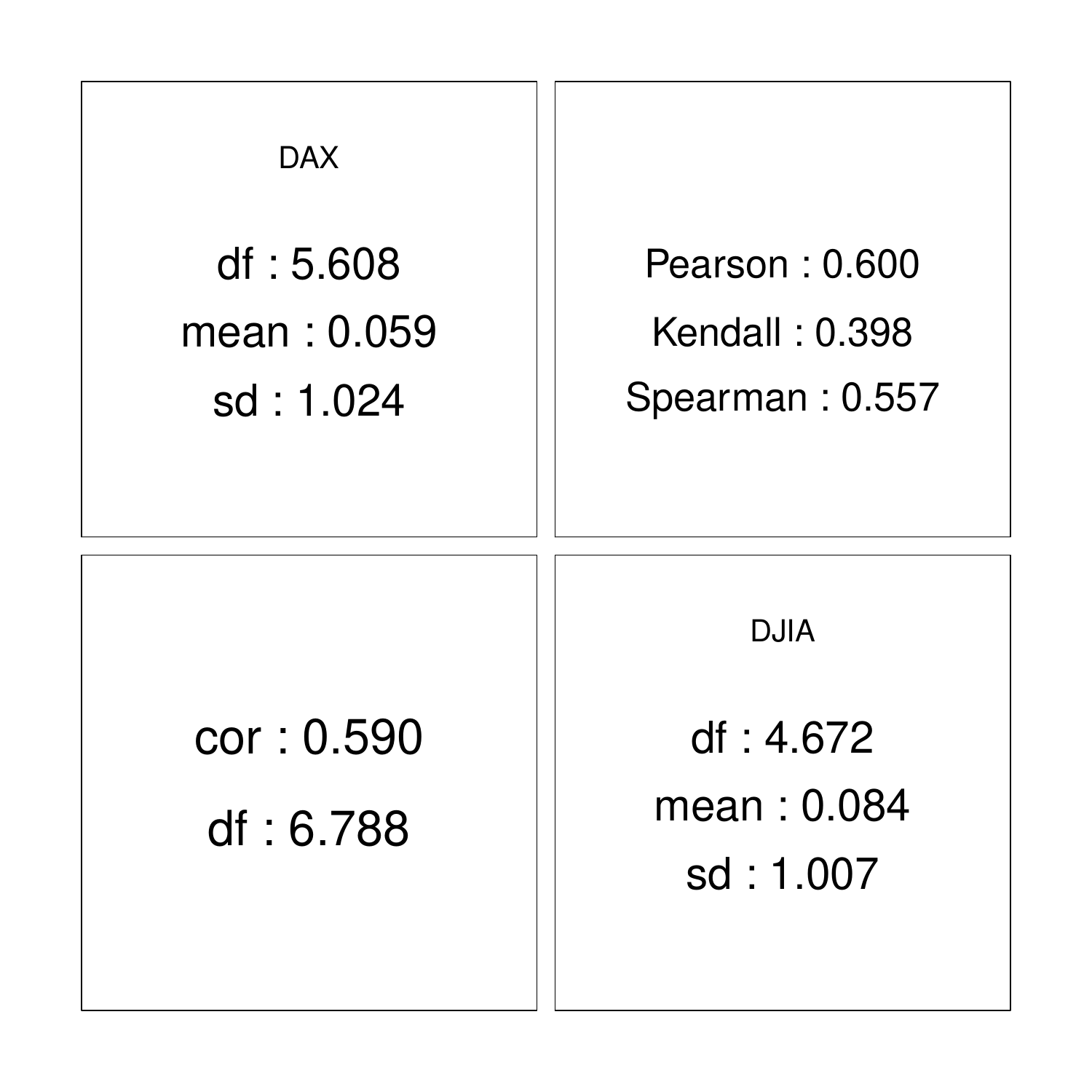}\\
		\label{fig:test2}
	\end{minipage}
	\caption{The innovation and copula distributions for the DAX and DJIA indexes.
		For the figure on the left, both diagonal panels show the folded cumulative distribution functions of the innovations (black line), together with a Normal distribution (dashed blue) and the fitted cdf of a Student distribution (dashed green).
		The top-right panel displays the innovations, the bottom-left panel the probtiles (i.e. the historical sample for the in-copula).
		The table on the right contains different measures related to the corresponding distributions on the left.
		On the diagonal, the estimated parameters of a Student distribution.
		The top-right panel gives different measures of dependency, the bottom-left panel the estimated parameters for a Student copula computed using a maximum-likelihood estimation.
	}
	\label{fig:innovationIN}
\end{figure}

The marginal distributions and the copula for our standard DAX-DJIA example is given in Figure \ref{fig:innovationIN}.
The bivariate distribution for the innovations is shown on the upper-right panel and has already been displayed in Fig.~\ref{fig:datasets}.
The diagonal panels show the folded cdf, namely $\cdf(\varepsilon)$ when $\cdf <= 0.5$, and $1 - \cdf(\varepsilon)$ when $\cdf > 0.5$.
The cdf can be computed simply by ordering and counting on an empirical sample, whereas a pdf evaluation needs more sophisticated algorithms involving a kernel which distorts/smoothes the resulting curve.
Then, folding the cdf allows to display both sides on a logarithmic vertical scale, showing on the same plot the core and both tails of the distribution.
The comparison with the normal and Student distributions shows clearly the up-down asymmetry for the innovations, with the down moves close to a Student distribution while the up moves show less extreme moves and a distribution  closer to a normal.
Such asymmetry is characteristic of stock indexes.
Notice also that the asymmetry becomes clear above 2 standard deviations, namely for large movements.

The bottom-left panel shows the corresponding in-copula.
The experts in copulas will notice the clear dependency along the diagonal, the symmetries around both diagonals and a dot density visually compatible with a Student copula (but not with a normal copula).
The right set of panels gives some numerical values for statistical estimations on the corresponding left panels.
Interestingly, the fits with a univariate marginal pdf and the bivariate copula all point to Student distributions with a small number of degrees of freedom (``df'' in the right figure).

The next subject is to statistically check if the empirical bivariate points are indeed compatible with the estimated copula.
This is done in two steps.
First, a Rosenblatt transform is used to transform the empirical copula to an independent copula.
Second, two statistical tests are used to check if the transformed copula is compatible with an independent copula.
Both points are the subject of the next two sub-sections.
The same strategy will be used for the out-copula in the next section, albeit using a forecast for the pdf and copula.

\subsection{The Rosenblatt transform}

Given an analytical copula $C$ and a set of empirical points $(z_1, z_2)$, the Rosenblatt transform uses $C$ to produce a new set of points $(u_1, u_2)$.
If the points $(z_1, z_2)$ are indeed distributed according to $C$, then the transformed points $(u_1, u_2)$ are distributed according to an independent copula, namely $C(u_1, u_2) =  \Pi(u_1, u_2) = u_1 \cdot u_2$.
In other words, the density of points for $(u_1, u_2)$ should be uniform on the unit square.
Essentially, this is a multivariate generalization of the PIT transform.

\begin{figure}[ht]
	\centering
	\begin{minipage}{.5\textwidth}
		\centering
		\includegraphics[width=0.9\linewidth]{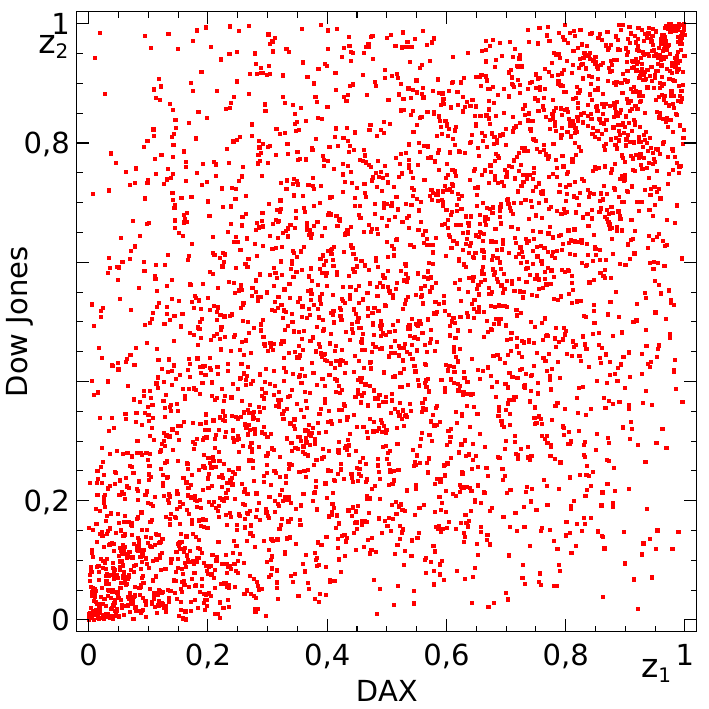} \\
	\end{minipage}%
	\begin{minipage}{.5\textwidth}
		\centering
		\includegraphics[width=0.9\linewidth]{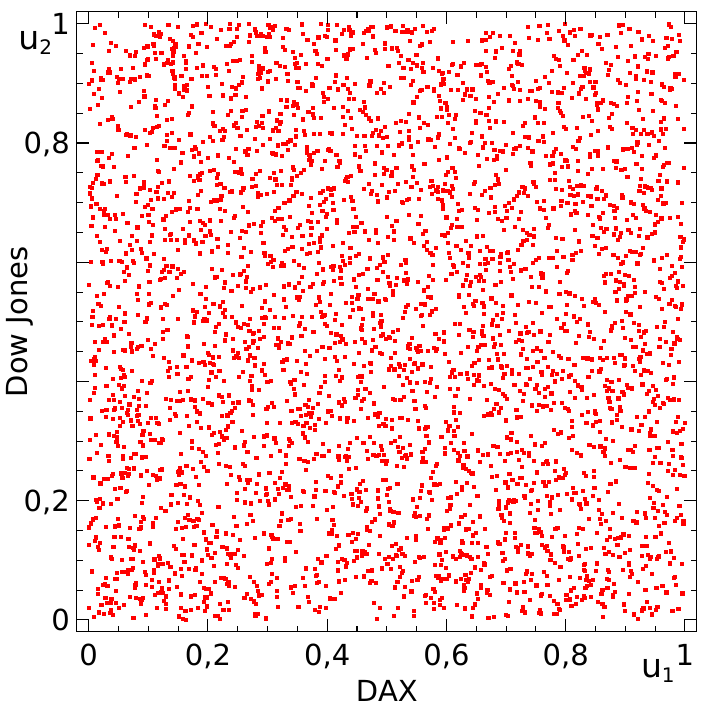} \\
	\end{minipage}
	\caption{The plots of in-sample probtiles  for $z$ (on the left) and the in-sample probtiles for $u$ after a Rosenblatt transform (right), called in-trf-copula.
	Visually, the transformed data looks uniformly distributed over the unit square, despite the $\sim$60\% correlation in the original sample.
		\label{fig:probtileRosenblatt-IN}}
\end{figure}

The Rosenblatt transform is defined in \citep{Rosenblatt.1952}.
In two-dimensions, it takes the form
\begin{subequations}
	\begin{align}
	u_1 & = z_1 \\
	u_2 & = P(z_2 \, | \, z_1)
	\end{align}
\end{subequations}
where $P(z_2 \, | \, z_1)$ is the conditional cumulative distribution function of $z_2$ knowing $z_1$.
The copula for $(u_1, u_2)$ will be called \textit{in-trf-copula}.
From the postulated analytical distribution of the copula (usually Student or Normal), it is possible to determine a closed form for the conditional cdf.
In our case, an analytical Student copula is used, with the parameters taken from the in-sample estimation.
The figure \ref{fig:probtileRosenblatt-IN} shows the transformation at work for the DAX-DJIA data set.

\subsection{Testing for independence}
\label{sec:testsIN}

A few statistical tests allow to check whether an empirical sample is distributed according to an independent copula.
Two different tests are used in this paper.

The first test is based on the work of \citep{GenestRemillard.2004} and \citep{GenestFavreBeliveauJacques.2007}, essentially computing a $L_2$ distance between the empirical copula and the independent copula.
Considering empirical data of length $n$, the empirical copula $C_n(u)$ can be evaluated.
This test is based on the test-statistic
\begin{equation}
	S_n^\Pi = \int_{[0,1]^d} n(C_n(u) - \Pi(u))^2 du
\end{equation}
where $\Pi(u)$ is the independent copula and $d$ the dimension of the copulas (this paper only consider the case $d = 2$).

As explained in \citep{GenestRemillard.2004}, a large sample of independent realizations (Monte-Carlo simulations) of this test statistic is computed $S_n^{\Pi,(1)},...,S_n^{\Pi,(N)}$ for a large $N$ under the null hypothesis of independence. An approximated $p$-value is then computed using
\begin{equation}
	p_\text{GR}(S) = \frac{1}{N+1} \left( \sum_{k = 1} 1(S_n^{\Pi,(k)} \geq S) + \frac{1}{2} \right).
\end{equation}
This test is called \textit{GR-Test} thereafter and is directly available in the package ``copula'' in R, see \citep{HofertKojadinovicMaechlerYan.2016}.

The second test is based on checking the number of empirical points on a regular tiling over the unit square  \citep{Knuth.1998}.
The unit square is divided in $N^2$ tiles of equal size, and the number of empirical points $n_i$ falling in each tile $i$ is counted.
The test-statistic is
\begin{equation}
	\label{def:sigmaTile}
	\sigma_\text{tile} = \sqrt{\frac{1}{N^2} \sum_{i} (n_i - \mu)^2}
\end{equation}
where $\mu = n/N^2$ is the expected number of points in one tile for a sample of size $n$.
We used a tiling with $N = 10$ for this paper, corresponding to roughly 38 empirical points per tile for our DAX-DJIA example.

This test statistic measures the standard deviation of the points' count.
The distribution of the test-statistics can be computed analytically for $\sigma_\text{tile}^2$ in an in-sample setting, but this is not possible in the out-of-sample setting.
For this reason, and because it is more natural to take  $\sigma_\text{tile}$ as the test variable, we use for both cases Monte Carlo simulations to compute the $p$-value function, similarly to the GR-test.
The test-statistic $\sigma_\text{tile}$ is computed for uniformly simulated data on $[0,1]^2$ (equivalent to an independent copula).
Repeating the computation with uniform samples, the cumulative density $\cdf_\text{MC}(\sigma_\text{tile})$ is obtained.
With empirical data, the tile-test statistic is denoted by $\sigma_\text{tile, emp}$ and the corresponding $p$-value is
\begin{equation}
	p_\text{tile} = 1 - \cdf_\text{MC}(\sigma_\text{tile, emp}).
\end{equation}

For both tests, the $p$-values give the probability that a value as large as $S_n^\Pi$ or $\sigma_\text{tile, emp}$ is obtained by chance.
The hypothesis of a uniform distribution can be rejected when this probability is too small, typically at the 5\% level.

\begin{figure}[ht]
	\centering
	\hspace*{-0.4cm}
	\begin{minipage}{.5\textwidth}
		\centering
		\includegraphics[width=1\linewidth]{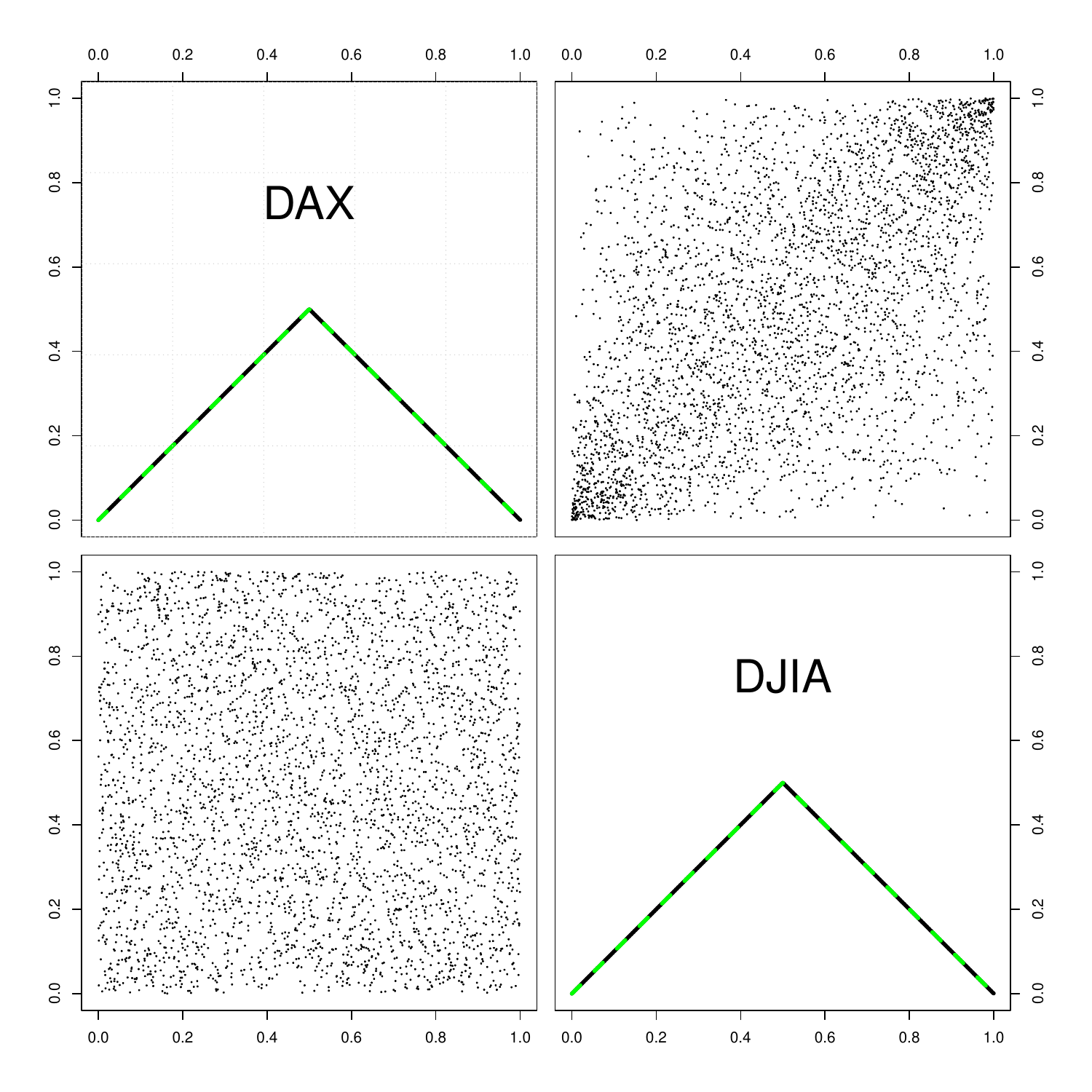} \\
	\end{minipage}%
	\hspace*{0.5cm}
	\begin{minipage}{.45\textwidth}
		\centering
		\includegraphics[width=1\linewidth]{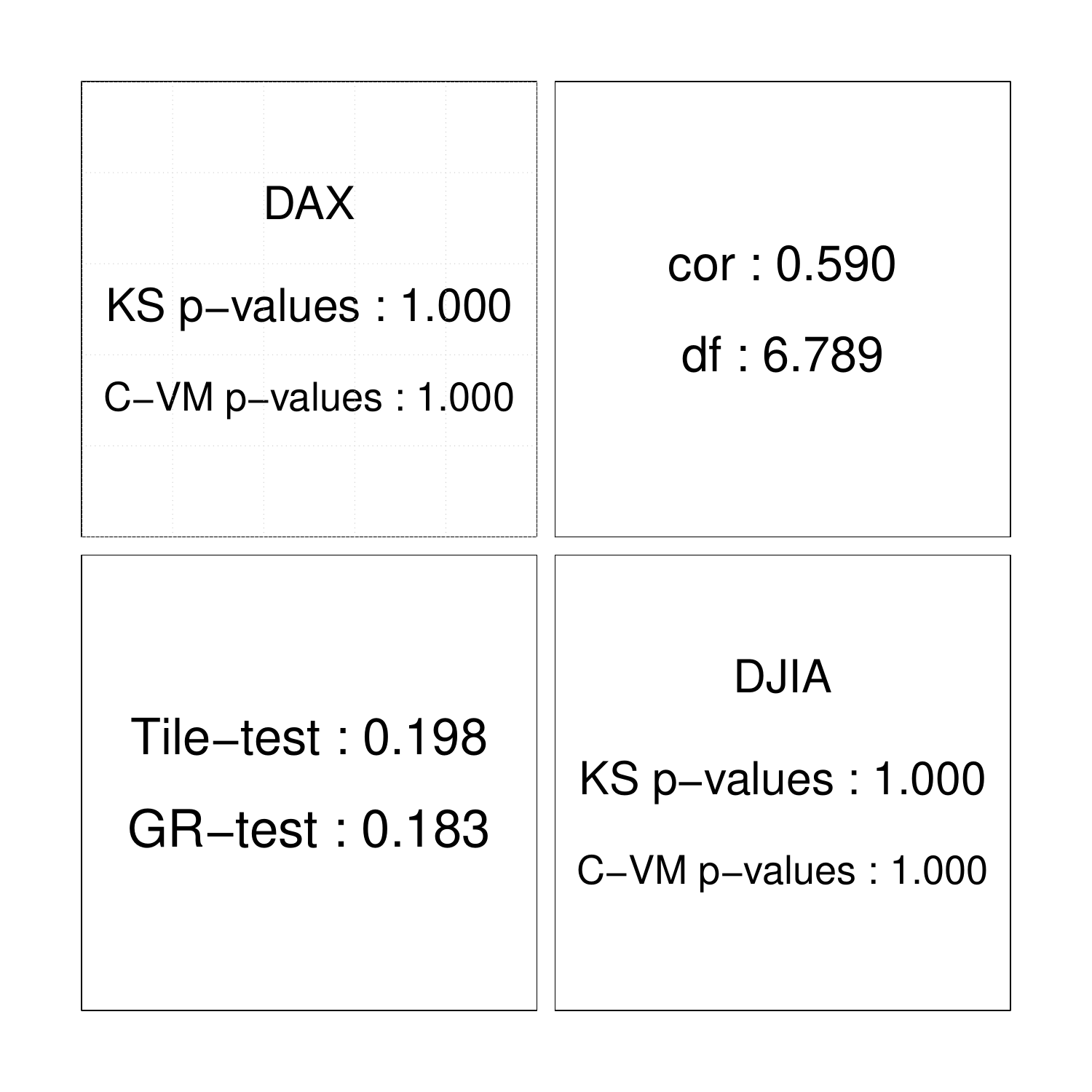}\\
	\end{minipage}
	\caption{
		The in-sample results for the DAX-DJIA example, with Rosenblatt transform and statistical tests.
		The set of panels on the left shows the empirical data.
		The diagonal shows the folded cumulative distribution functions of the probtiles (black), with the cdf of a uniform distribution over $[0,1]$ (green).
		Since computed from the empirical sample, the marginal distributions are exactly uniform.
		The top-right panel shows the in-copula, the bottom-left panel the in-trf-copula.
		The set of panel on the right contains estimated parameters and $p$-values.
		On the diagonal, the test for uniformity using the Kolmogorov-Smirnov and Cram\'er-von Mises tests are reported, with the corresponding $p$-values.
		By construction, these $p$-values are 1 in sample, denoting a perfect uniformity, but it is not the case out-of-sample (see Fig.~\ref{fig:probtilesOUT}).
		In the top-right panel, the parameters of a Student copula fit is reported, and on the bottom-left, the $p$-values for independence tests of the in-trf-copula.}
	\label{fig:probtilesIN}
\end{figure}
The results of the Rosenblatt transform and the corresponding statistical tests is conveniently summarized in
figure~\ref{fig:probtilesIN} for our DAX-DJIA example.
In this example, the test does not reject the hypothesis of independence for the in-trf-copula at a 5\% level, hence does not reject using a Student copula to describe the distribution of probtiles.

In order to summarize the computations, the sequence of transformations that have been used are
\begin{equation}
  r~~ \stackrel{1}{\longrightarrow} ~~\varepsilon~~\stackrel{2}{\longrightarrow} ~~z~~ \stackrel{3}{\longrightarrow} ~~u~~ \stackrel{4}{\longrightarrow} ~~p\text{-value}
\end{equation}
where the transformation 1 uses a volatility forecast to remove the heteroskedasticity in order to have a stationary problem, 2 is a PIT that removes the marginal distributions to get uniform random variables, 3 is a Rosenblatt transform that maps the empirical copula to a uniform copula (if the chosen model copula describes correctly the empirical data), and 4 is a statistical test checking the uniformity of the copula for $u$.
The out-of-sample setting will use the same sequence of mappings, with the added complexity of tracking the out-of-sample information associated to a forecast.

\FloatBarrier
\section{In-sample empirical results}
\label{sec:inSampleResults}

\subsection{Set-up and computational details}
The computations described in the previous section have been applied on a large set of major indexes and FX rates.
The indexes have been selected in order to cover the main asset classes (stock markets, fixed incomes, commodities) and to contain some negative correlations, for a total of 21 indexes.
The sample is completed by 6 major FX rates.

Depending on the time series, the start date is 1998-10-02, the end date ranges from 2021-03-04 to 2021-05-13.
All the assets are tested pairwise, for a total of $351$ unordered pairs or $702$ ordered pairs (the order matters in the Rosenblatt transform).
Each pair is synchronized on the available daily innovations, discarding dates with 0 or 1 values.
The resulting data sets have a length ranging from $5\,311$ to $5\,900$ points.
This large bivariate empirical set allows for interesting cross-sectional studies, presented in this section.

Due to holidays and zero returns (induced by the price finite granularity), some cares need to be taken when computing the cdf.
After volatility discounting, the zero returns are mapped to zero innovations.
An excess of zero innovations induces a step in its cdf, and all zero innovations are mapped to one $z$ value, depending on the algorithm (often corresponding to the left limit $\cdf_\text{inf}$ for the cdf at the step).
The distribution for $z$ is then not uniform, resulting in K-S uniformity test with $p$-values lower than 1.
Also, the in-copula and in-trf-copula show an excess of points along lines in the middle of the unit square, resulting in the uniformity tests to fail.
The solution is to draw randomly $z$ in the step, namely to draw a value between  $\cdf_\text{inf}$ and $\cdf_\text{sup}$ with a uniform probability.
This procedure is known as ``randomly breaking the ties'', and in the R-package \texttt{copula}, this can be done using an option in the \texttt{pobs} function.
With this modification, the empirical univariate distributions become perfectly uniform and consistent with the theoretical setting.

\subsection{Estimated Student copulas}
\begin{figure}[ht]
	\centering
	\includegraphics[width=1\linewidth]{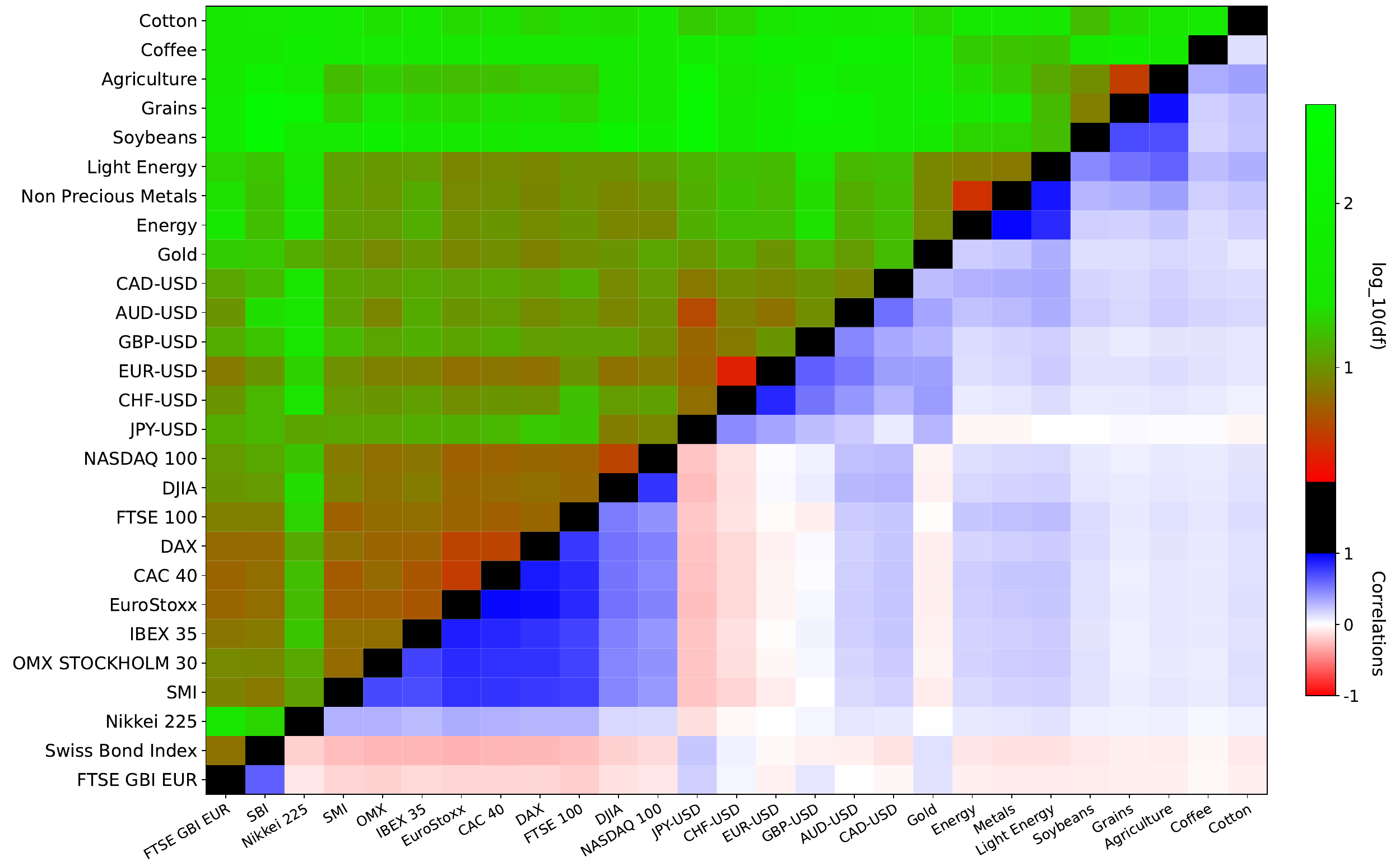}
	\caption{
		The correlations (bottom-right) and log-degrees of freedom (upper-left) of an estimated bivariate Student in-sample copula for each pair of assets.
		The $\log(df)$ are described in the top-left part of the graph with colors varying from red to green as explained in the top part of the color bar.
		The correlations are displayed in the bottom-left section of the graph, with colors varying from red to blue. }
	\label{fig:df_corr_in}
\end{figure}

\begin{figure}[ht]
	\centering
	\includegraphics[width=0.8\linewidth]{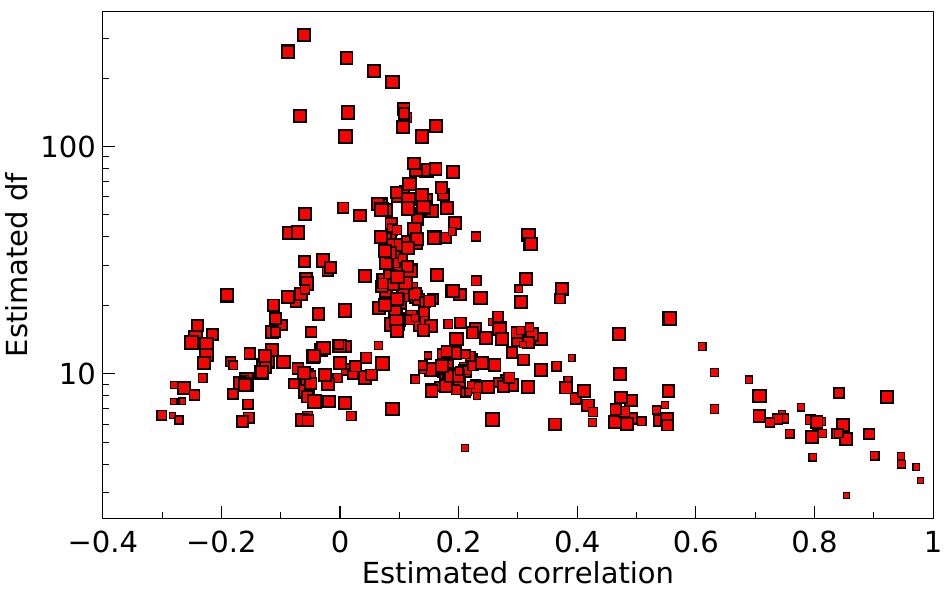}
	\caption{
		Cross-section of the estimated parameters of the Student copula on 351 empirical pairs of assets.
		The sizes of the symbols are proportional to the $p$-values for the tile-test (more precisely to 1 + $p$-value).}
	\label{fig:df_vs_corr}
\end{figure}

The first step in the empirical analysis is to estimate a Student copula for each pair of probtiles.
Figures \ref{fig:df_corr_in} and \ref{fig:df_vs_corr} show the estimated parameters of the pair-wise Student copulas, namely correlations and number of degrees of freedom $df$.
For the figure \ref{fig:df_corr_in}, the time series have been ordered so as to minimize the off-diagonal correlations, namely permutations are done between the series to minimize the sum over the correlation matrix of $(1+\log(d_{i,j}))|\rho_{i, j}|$, with $d_{i,j}$ being the distance to the diagonal with periodic boundary conditions and $\rho_{i, j}$ the correlation.
The ordering has been done on the in-sample correlations obtained from the copula estimates, the same ordering is used out-of-sample.

The estimation procedure is symmetric in the exchange of both time series.
This symmetry is used in figure \ref{fig:df_corr_in} to represent both the correlation and $df$ in the same graph.
The chosen representation makes clearly visible the inverse relation between estimated correlation and $df$ in the cross-section.
In the small correlation domain (and $df$ not too small), it is increasingly difficult to make the difference between Student, normal and independent copulas.
Consequently, $df$ cannot be estimated reliably, leading to large values (i.e. close to a normal copula).
For large correlations, say $|\rho| > $ 0.25, $df$ tends to decrease with increasing $\rho$, but the $p$-values tend also to be smaller.

Another representation of the same data is given on Fig.~\ref{fig:df_vs_corr}, where the size of the symbols is proportional to the $p$-values for the tile-test (see the next subsection).
As could be expected, the fit by a Student copula tends to be weaker for high correlations and low $df$.
This is the domain of parameters where the Student copula shows best its characteristics, and where the test with empirical data is the most discriminating.

In order to test for a spurious effect induced by the estimation of the Student copula, a Monte Carlo simulation has been done as follows.
Random samples are drawn according to a bivariate Student distribution, with fixed degrees of freedom $df = 6$ and increasing correlations in the range $[0.1, 0.95]$, and with the same length as the index sample.
These random samples are estimated for a Student copula, with the resulting $\rho$ and $df$.
A simple plot shows that $df$ is independent of $\rho$, and with a standard deviation of the order of $0.7$.
This shows that the decrease of $df$ with increasing $\rho$ in Fig.~\ref{fig:df_vs_corr} is outside the standard deviation and not due to the estimation procedure.

This relation between correlation and $df$ is new.
It shows that assets with large correlations tend to have stronger co-movements.
The origin is likely to be found in the trading of strongly dependent assets and in the (over-)reactions of the traders to moves in correlated pairs.

Implicitly, the in-sample test assumes that the parameters are constant in time, in particular that the correlations between innovations stay constant.
At a scale of more than 20 years, this is not an obvious assumption, particularly for large correlations.
The instability in time of the parameters could be one cause for the lower $p$-values.
A study in this direction would require longer time series with several decades of data in order to have significant results.

Turning to the quest for a multivariate data generating process, a complete multivariate model for these assets can be constructed with Student marginal distributions and copulas.
In details, the present results point to a complex model, with $df$ to be specified for each marginal and copulas, together with the correlation matrix.
Besides, in order to be complete, the present pair-wise study must be extended to the full n-dimensional copula.
A drastic simplification is to assume one value for $df$, for all the marginals and copulas, say for example $df = 6$.
The model for the multivariate innovations is then a Student distribution with the specified correlation matrix and the selected $df$.
In light of the weak dependency of $df$ on $\rho$ and of the good overall fit provided by Student copulas, such a model would be a good multivariate description of the financial time series for the innovations (and certainly better than a multivariate normal for the returns).

\subsection{$p$-values for the Student copulas}
\label{subsec:p-Values}
The second step in the empirical study is to use the in-trf-copula to check the adequacy of the Student assumption, using both the Genest-R\'{e}millard and tile-tests.
For both tests, the $p$-value evaluation involves the generation of the cumulative distribution functions by Monte Carlo simulations.
One key parameter for the computation is the number of points $n$ in the empirical bivariate copula.
Since this computation is rather costly, the Monte Carlo simulations are done for the smallest and the largest lengths among the pairs of data, respectively for $5\,311$ and $5\,900$ points.
For an actual length $n$, the $p$-values are obtained by linear interpolation in $n$ between these 2 sample sizes.

\begin{figure}[ht]
	\centering
	\includegraphics[width=1\linewidth]{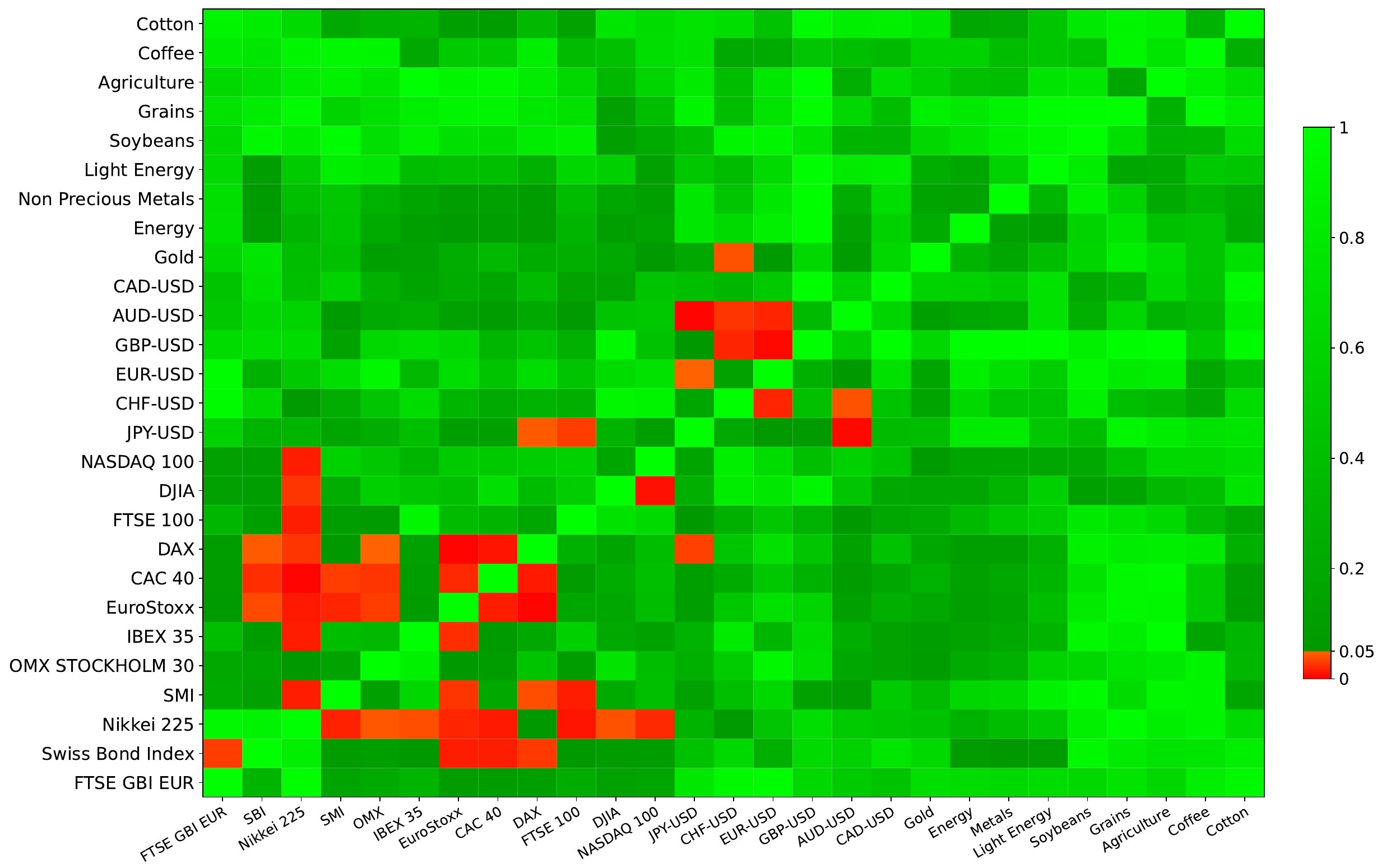}
	\caption{
		$p$-values of the GR-test performed in-sample for each pair of assets.
		Red (resp. green) entries correspond to (resp. non-)rejection of the uniformity test at the $5\%$ level.
		The Rosenblatt transform is made for each pair of variables, and the x-axis corresponds to the first variable.
	}
	\label{fig:inGR}
\end{figure}

\begin{figure}[ht]
	\centering
	\includegraphics[width=1\linewidth]{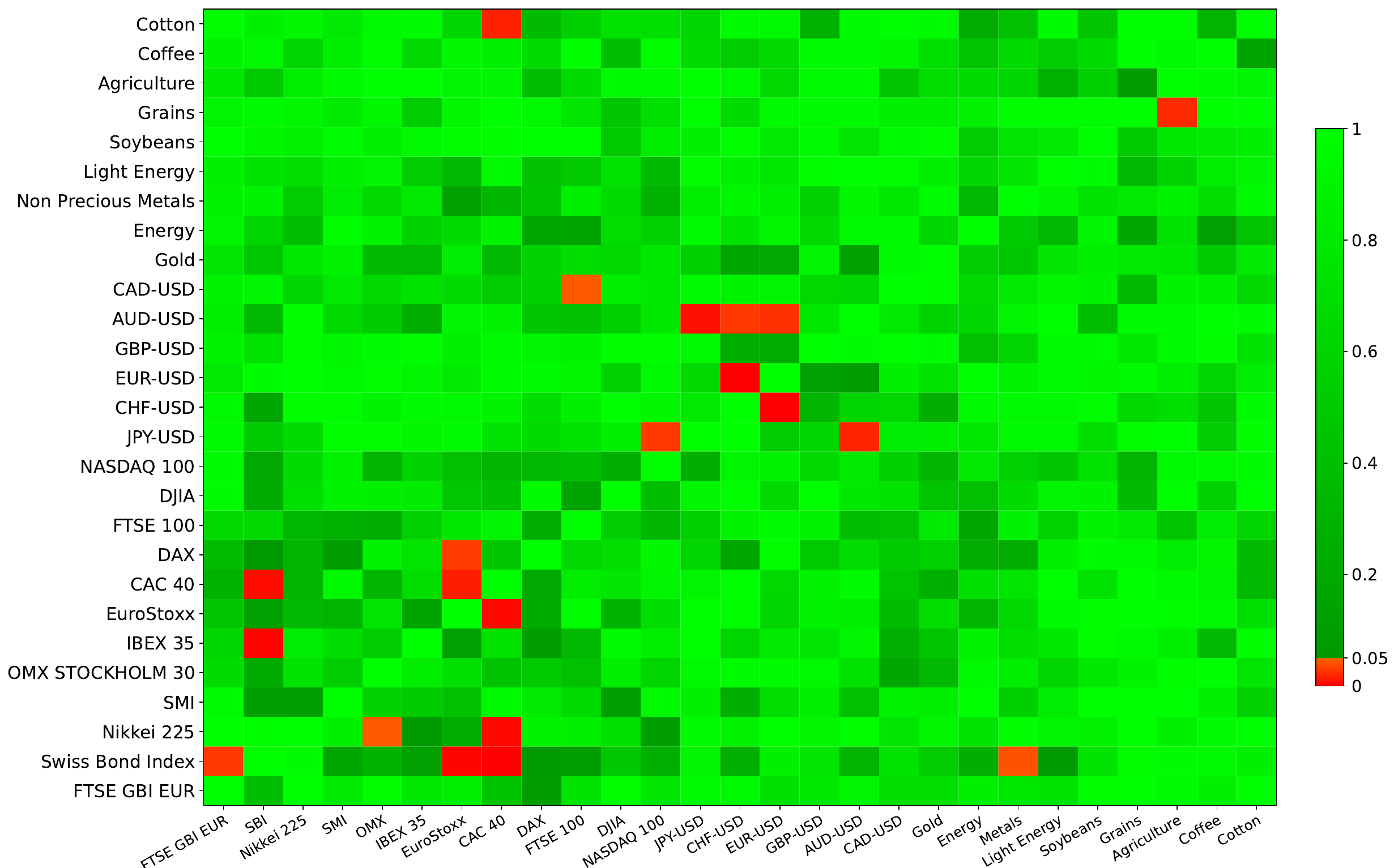}
	\caption{
				$p$-values of the tile-test performed in-sample for each pair of assets.
				The layout is identical to figure~\ref{fig:inGR}. }
	\label{fig:inTile}
\end{figure}

The $p$-values for both independence tests are given in figure \ref{fig:inGR} for the GR-test and in figure \ref{fig:inTile} for the tile-test.
The Rosenblatt transform is not symmetric under the exchange of time series, and the full matrix of $p$-values is plotted.
Overall, the $p$-values show mostly non-rejection of the null hypothesis of uniformity of the in-trf-copula, validating a Student copula to describe the dependencies between financial time series.

The $p$-values for the GR-test are mostly lower than for the tile-test.
A cross-section of $p$-values for the GR-test and tile-test shows a positive but weak relation between them, indicating that they are sensitive to different non-uniform deviations of the copula.
Since visually the empirical copulas look very uniform, it is difficult to pinpoint the causes for the differences, and further theoretical and numerical studies would be required.

Some uniformity tests are rejected.
For a perfect model, we can expect 5\% of the tests to be randomly rejected, the actual rates are 52/702 = 7.4\% for the GR-test and 21/702 = 3.0\% for the tile-tests.
Hence, the rejection rates are close to the expected value.

Yet, most rejection cases are between time series with a large correlation, mainly stock indexes.
Using $n_i - \mu$ from Eq.~\ref{def:sigmaTile}, some detailed plots of the deviations from a uniform density of points show random tiles with large deviations, but no particular structures seems to emerge.
A peculiar example of rejection is the FX couple CHF-USD and EUR-USD, which is strongly correlated with $\rho$ = 81\%.
A detailed plot of the deviations from a uniform density shows a structure across the unit square: significant deviations are not just spread randomly on the square, but form roughly a 30 degree line.
Specific to this time series pairs, the CHF had a cap against the EUR enforced by the Swiss national bank (SNB) during the period from 6 September 2011 to 15 January 2015.
To enforce this policy, the SNB had to buy large amounts of currencies, possibly altering the dependencies with the EUR.
In order to check for the impact of the cap, the same study was done on the period January 1, 1990, to January 1, 2001, during which the CHF was free floating.
Essentially the same results are obtained, with similar deviations from uniformity.
Hence, it is possible that some smaller substructures should be introduced above a core Student copula for this pair.

The overall conclusion of this large cross-sectional study is that Student copulas seem adequate to describe the bivariate dependencies between time series of innovations (more precisely, is not rejected in the majority of cases).
The estimated number of degrees of freedom $df$ is slowly decreasing with increasing correlation.
Moreover, a much simpler model with $df \simeq 6$ is a good multivariate approximation.
Because of the good results with a Student copula, we have not tried other copulas.
Possibly, an even better description of the empirical data could be obtained, say for example with the copula generated by a multivariate \textit{non-central} Student that would account for the up-down asymmetry in the marginal observed in Fig.~\ref{fig:innovationIN}.

\FloatBarrier
\section{Risk and copula: out-of-sample backtest}
\label{sec:out-of_sample}

\subsection{The out-of-sample set-up}
\label{subsec:outsetup}
Risk evaluation is fundamentally a forecast for the possible (large) losses.
It is a ex-ante computation made at a time $t$ for a risk horizon $\DT$, producing a forecast for the probability distribution  $\tilde{p}_{t, \DT}(r)$  of the returns at time $t+\DT$.
From this distribution, the standard risk measures can be computed, as value-at-risk (VaR) and expected shortfall (ES or CVaR).
After a time interval $\DT$, one realized return $r[\DT](t+\DT)$ is obtained, an ex-post value.
In principle, $r$ should be drawn from $\tilde{p}(r)$, but $\tilde{p}$ is time dependent.
The fundamental problem of backtest is to assert if the sequence of returns $r(t)$ is a sequence of one draw from $\tilde{p}_t(r)$.
In this form, it is fundamentally a non-stationary problem.
Our goal is to investigate the risk forecast backtest in a bivariate set-up, where $r$ is a 2-dimensional vector, and $\tilde{p}$ the corresponding probability.

The core ideas to map this problem into a tractable one are to use univariate PIT on the stationary innovations to obtain a copula, then to use a Rosenblatt transform to have a uniform copula.
Yet, both mappings should be done while respecting the ex-ante/ex-post information, a fundamental difference from the in-sample computation.
The out-of-sample transformed copula can then be tested for uniformity, using an adapted version of the tile-test presented in the previous section.
The key difference from the in-sample set-up is that such risk forecasts are based on a sampling, and not on an analytical formula for the multivariate pdf, say a multivariate Student.

The relevant aspects of the risk computation are presented here and are based on a historical innovations methodology in order to account correctly of the heteroskedasticity.
The risk horizon is denoted with $\DT$.
The empirical study is done with $\DT = 1$ business day in order to maximize the number of independent points in the empirical samples.
The computation of the innovations $\varepsilon$ has been presented in Sec.~\ref{sec:stationarity}.
At time $t$, using historical prices and formula~\eqref{eq:innovation}, a sample of 500 historical innovations $\varepsilon(t-i)$ is produced.
A set of 500 future scenarios for the returns is computed with
\begin{equation}
	r^{(i)} = \tilde{\sigma}(t) \,\varepsilon(t-i) \hspace{3em} \text{for~~~} \; 0 \leq i <  500 \label{eq:returnScenarios}
\end{equation}
where $\tilde{\sigma}(t)$ is a one-day forecast for the volatility, computed at time $t$ with the previous returns, using the long memory ARCH weights.
The set of 500 scenarios $r^{(i)}$ is indexed by the time lag of the corresponding innovations $\varepsilon(t - i)$.
This set of scenarios spans the probability distribution $\tilde{p}_{t, \DT}(r)$ of the returns.
Let us emphasize that the pdf is based on an empirical sample and not on an analytical formula.

In a multivariate context, these equations are applied independently on each time series.
The corresponding multivariate ARCH process is a simple independent cross-product of univariate ARCH processes, with the same long-memory volatility model for each component.
In particular, there is no cross-product term in this model.
Yet, the synchronicity is important in order to keep the multivariate dependencies between time series.
For example, the bivariate forecast is produced by the vector $(r_1^{(i)}, r_2^{(i)})$, and similarly for larger dimensions.

At the time $t+\DT$, the ex-post return is obtained.
For each time series, the corresponding innovation is
\begin{equation}
	\varepsilon(t+\DT) = \frac{r[\DT](t+\DT)}{\tilde{\sigma}(t)}   \label{eq:def:exPostInnovation}
\end{equation}
In equations \eqref{eq:returnScenarios} and \eqref{eq:def:exPostInnovation}, the same volatility forecast at time $t$ is used and is just a multiplicative factor.
Hence, it can be simplified, and the forecast for the distribution of the innovations $\tilde{p}_{t, \DT}(\varepsilon)$ can be compared with the realized innovation $\varepsilon(t+\DT)$.
As the multivariate model is a simple cross-product, the same simplification can be done in a multivariate context.
This simplifies the backtest to studying the multivariate innovations only and more precisely the probability forecasts and realized innovations.
For risk forecasts with another structure, the returns must be taken as the base variables, but the principle is identical.

In the in-sample set-up, the PIT is computed using the full sample.
The principle of the transformation is similar for the out-of-sample set-up, but the pdf forecast is used at each time $t$
\begin{equation}
	\label{eq:probtileOutOfSample}
	z(t+\DT) = \tilde{P}_{t, \DT}\left(\varepsilon(t + \DT)\right)
\end{equation}
where $\tilde{P}_{t, \DT}$ is the cdf computed at time $t$ for a risk horizon $\DT$.
For a risk horizon of 1 business day, if the sequence of $\varepsilon(t + 1d)$ is drawn from $\tilde{P}_{t, 1d}$, then the distribution of $z$ should be iid with a uniform distribution.
Notice that in this set-up, this is a non-trivial statement since $z$ is computed out-of-sample, in particular it should be checked that $z$ has a distribution close enough to uniform and is independent.
A new statistical test has been presented in \citep{Zumbach.TileTest} in order to assess the validity of this statement, we will take for granted in this paper that this is the case, or at least that this is close enough to be the case.
Indeed, if the univariate pdf is not close enough to uniform, the bivariate copula test below will fail.

In a bivariate case and for a risk horizon of 1 day, a sampling of the \textit{out-copula} is obtained by $(z_1(t+1d), z_2(t+1d))$, a sampling from $\tilde{P}_{t, 1d}(\varepsilon_1, \varepsilon_2)$ at the point $(\varepsilon_1(t+1d), \varepsilon_2(t+1d))$ at each time $t$.
Since $\varepsilon$ and $z$ should have stationary distributions, it can be expected that the in-copula and out-copula for $z$ have similar characteristics.
Yet, our goal is not to characterize the out-copula, but to assert if the sequence $(\varepsilon_1(t+1d), \varepsilon_2(t+1d))$ is originating from the forecast $\tilde{P}_{t}(\varepsilon_1, \varepsilon_2)$, where $\tilde{P}_{t}$ is constructed here from a sample of 500 past historical innovations.
The strategy is the same as in the in-sample setting, namely using a Rosenblatt transform to map to a copula that must be uniform if the forecasts and realizations agree and to test for uniformity.

\subsection{The Rosenblatt transform out-of-sample}

In the in-sample study in Sec.~\ref{sec:inSample}, the full empirical data set is used to estimate an analytical form for the copula, then the analytical form is used in the Rosenblatt transform to compute the conditional expectations.
In the out-of-sample setting, the probability forecast is spanned by a set of 500 data points (at each time $t$), and we want to use directly this sample to compute the transform, without estimating an analytical copula.

In the Rosenblatt transform, the idea is to use a Gaussian kernel on the historical data points to compute the conditional expectations.
At time $t$, the in-sample probtiles are $(z_1^{(i)},z_2^{(i)})$ for $i = 0,...,499$, and $(z_1, z_2)$ are the out-of-sample probtiles, all computed from the sampled bivariate pdf $\tilde{P}_{t}(\varepsilon_1, \varepsilon_2)$.
The empirical Rosenblatt transform $(u_1,u_2)$ of the ex-ante probtile $(z_1,z_2)$ is defined by
\begin{subequations}
	\label{eq:empiricalrosenblatt}
\begin{align}
	u_1 & = z_1 \\
	u_2 & = \frac{ \sum_{i=0}^{499} w_i \,\mathbf{1}\left(z_2^{(i)} \leq z_2\right)}{\sum_{i} w_i} \hspace{2em}\text{with}~~~ w_i = w\left(z_1^{(i)} - z_1; \delta\right) \label{eq:empiricalrosenblatt_u2}
\end{align}
\end{subequations}
where  $w$ is a weight function with $\delta$ fixing the band width.
After some experimentation with various analytical forms for $w$, good results are obtained with Gaussian weights
\begin{equation}
w(r; \delta) = \exp\left (-\frac{r^2}{2\delta^2}\right)
\end{equation}
with the band half-width $\delta = 0.03$.
This computation is repeated for all the times $t$ in the sample in order to obtain the sample copula for $u$.
Both variables can be exchanged in the definition, without noticeable changes on the resulting distribution.

\begin{figure}[ht]
\centering
\begin{minipage}{.5\textwidth}
	\centering
	\includegraphics[width=0.9\linewidth]{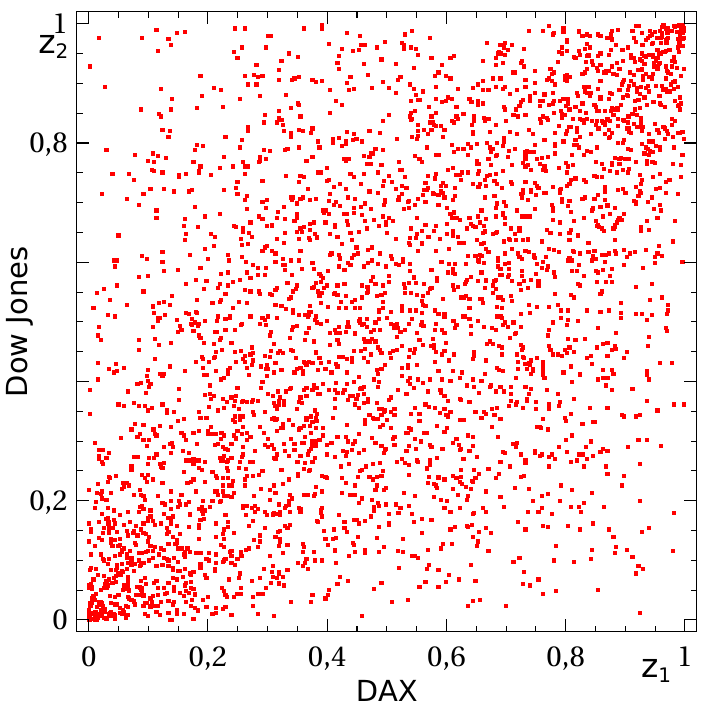} \\
\end{minipage}%
\begin{minipage}{.5\textwidth}
	\centering
	\includegraphics[width=0.9\linewidth]{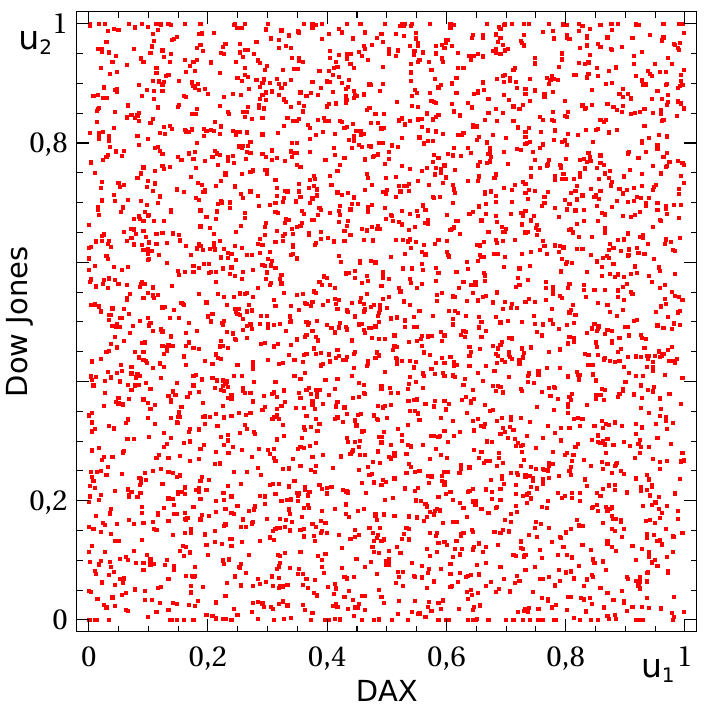} \\
\end{minipage}
\caption{The plots of the out-of-sample copula, for the probtiles on the left (out-copula) and after Rosenblatt transform on the right (out-trf-copula).
	\label{fig:probtilesrosOUT}
}
\end{figure}
To illustrate this empirical transformation for the DAX and DJIA, figure \ref{fig:probtilesrosOUT} shows the out-copula for $(z_1, z_2)$ and out-trf-copula for $(u_1, u_2)$ computed with the empirical Rosenblatt transform.
Notice the strong similarity with Fig.~\ref{fig:probtileRosenblatt-IN}.
The out-trf-copula on the right looks uniform, showing visually the effectiveness of the bivariate forecast.
This has to be formally tested, using the uniformity tests used in-sample.

\subsection{Testing for independence: out-of-sample case}

The right plots in figures \ref{fig:probtileRosenblatt-IN} and \ref{fig:probtilesrosOUT} look very similar, and a test for uniformity can be expected to give similar results.
This is not the case, the test values are clearly smaller in the out-of-sample case, leading to systematically larger $p$-values and to a broad acceptance of the forecasts.
The origin is fairly subtle, and lies in the empirical Rosenblatt transform.
For the sake of the explanation, let us take a rectangular kernel for $w(r)$, with the same half-width of 0.03.
With this set-up, the effective number of points in the sum in Eq.~\eqref{eq:empiricalrosenblatt_u2} is $2\cdot 0.03 \cdot 500 = 30$, and the point in the $u_2$ direction are spaced by 1/30 for a given $u_1$.
This effect spreads regularly the points in the $u_2$ direction, leading to a more uniform distribution compared to random uniform points.
Hence, the smaller values for the uniformity test.

In order to quantify the effect, Monte Carlo simulations are used in a set-up reproducing the out-of-sample setting for Student distributed data.
For a given correlation and number of degrees of freedom, an ordered sample of 500 random bivariate innovations are drawn according to a Student distribution.
The sample marginal cdf are computed using these 500 points.
One draw for the innovations is done for the out-of-sample point.
Using the marginal cdf, the probtiles $(z_1, z_2)$ of the out-of-sample innovations are obtained.
Similarly, the probtiles $(z_1^{(i)}, z_2^{(i)})$ corresponding to the 500 innovations are computed using their cdf.
Then, the empirical Rosenblatt transform is computed for the out-of-sample point using \eqref{eq:empiricalrosenblatt}, with the 500 points as the empirical sample for the copula.
This procedure gives 1 point for $(u_1, u_2)$, say for a given time $t$.
Then, the oldest in-sample point is dropped, the out-of-sample point is added in the sample, and a new out-of-sample point is drawn.
This procedure is repeated for the same length as the empirical sample, and the statistical uniformity tests $t$ (tile-test and GR-test) are computed on this sample for $u$.
This procedure is repeated a large number of times, and the cdf of the statistical test $t$ is obtained.
Finally, the $p$-value as function of the statistical test $t$ is $p(t) = 1 - \cdf(t)$.
This gives the probability $p$ to obtain a value as large as $t$ by chance, assuming data distributed according to a Student distribution for a given correlation and sample length and a forecast based on a sliding window of 500 points.
Finally, the dependency on the correlation needs to be investigated.

\begin{figure}[ht]
	\centering
	\includegraphics[width=0.8\linewidth]{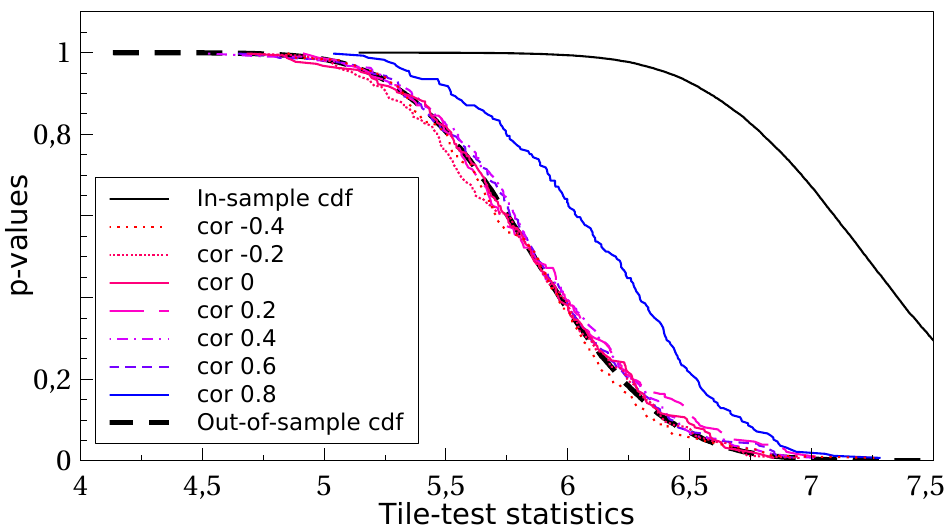}
	\caption{The $p$-value for the tile-test obtained by simulations for 5811 points.
		The samples are drawn according to a Student copula with 6 df.
		The in-sample case is plotted in black.
		The out-of-sample $p$-values are plotted in colors depending on the correlation.
	}
	\label{fig:tilecdfs}
\end{figure}
\begin{figure}[ht]
	\centering
	\includegraphics[width=0.8\linewidth]{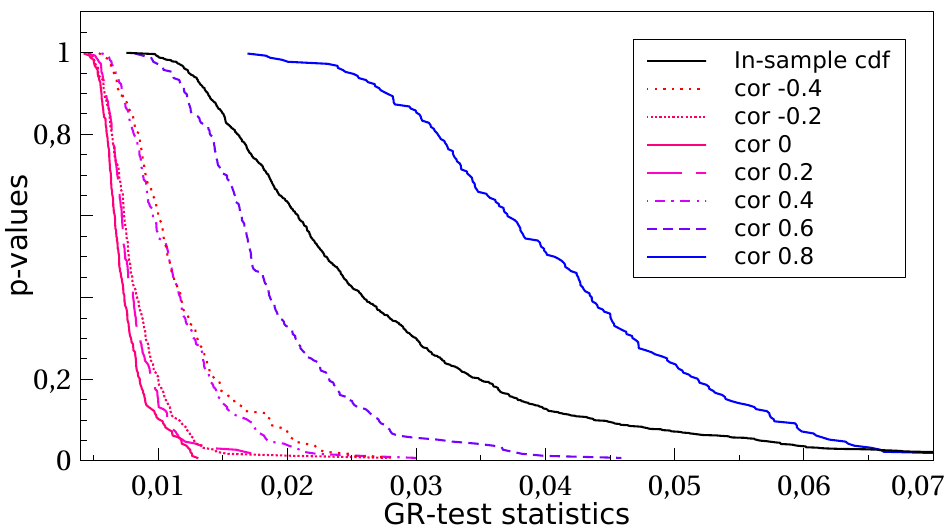}
	\caption{The cdfs for GR-test simulations for 5811 points, with the same setting as Fig.~\ref{fig:tilecdfs}. }
	\label{fig:grcdfs}
\end{figure}
For the relevant correlations, figures \ref{fig:tilecdfs} and \ref{fig:grcdfs} display $p(t)$ for the tile-test and the GR-test.
For both tests, the large shift to smaller values for the test statistics is very clear, and cannot be neglected.
For example with the tile-test, a value for $t = 6.5$ is around a 5\% rejection threshold, while it would be largely accepted using the in-sample $p$-value.

For the tile-test, the correlation of the innovations does not have much impact on the cdf, which appears as a simple shift compared to the in-sample setting.
We have used the simple ``shift and scale'' approximation to compute the out-of-sample $\cdf$ from the in-sample $\cdf$
\begin{equation}
		\cdf_\text{out-of-sample}(t) = \cdf_\text{in-sample} \left[\frac{q_{\text{in,}0.8} - q_{\text{in,}0.2}}{q_{\text{out,}0.8} - q_{\text{out,}0.2}}
		\left(t - q_{\text{out,}0.5} \right) + q_{\text{in,}0.5} \right]
	\label{eq:inToOutCdf}
\end{equation}
where $t$ is the tile-test computed in the out-of-sample setting, and $q_{\text{in,}\alpha}$ is the $\alpha$-quantile of the in-sample cdf, namely $q_{\alpha} = \cdf^{-1}(\alpha)$ for the respective cdf.
The reference out-of-sample copula used to compute $q_{\text{out,}\alpha}$ is the Student copula with correlation 0.4 and 6 $df$.
In the case of Fig \ref{fig:tilecdfs}, the simulations lead to a scale parameter : $\left(q_{\text{in,}0.8} - q_{\text{in,}0.2}\right)/\left(q_{\text{out,}0.8} - q_{\text{out,}0.2}\right)  = 1.21$.
Notice that with this simple approximation, the $p$-values for large correlations tend to be underestimated (the difference between the black-dashed curve and the blue curve for $\rho = 0.8$), leading to stronger rejections.

For the GR-test, the influence of the correlation cannot be ignored.
This is problematic because the $p$-values are depending on the distribution of the original data that are tested.
For this reason, we decided not to use the GR-test to assess independence in the out-of-sample case.

\begin{figure}[ht]
	\centering
	\begin{minipage}{.5\textwidth}
		\centering
		\includegraphics[width=1\linewidth]{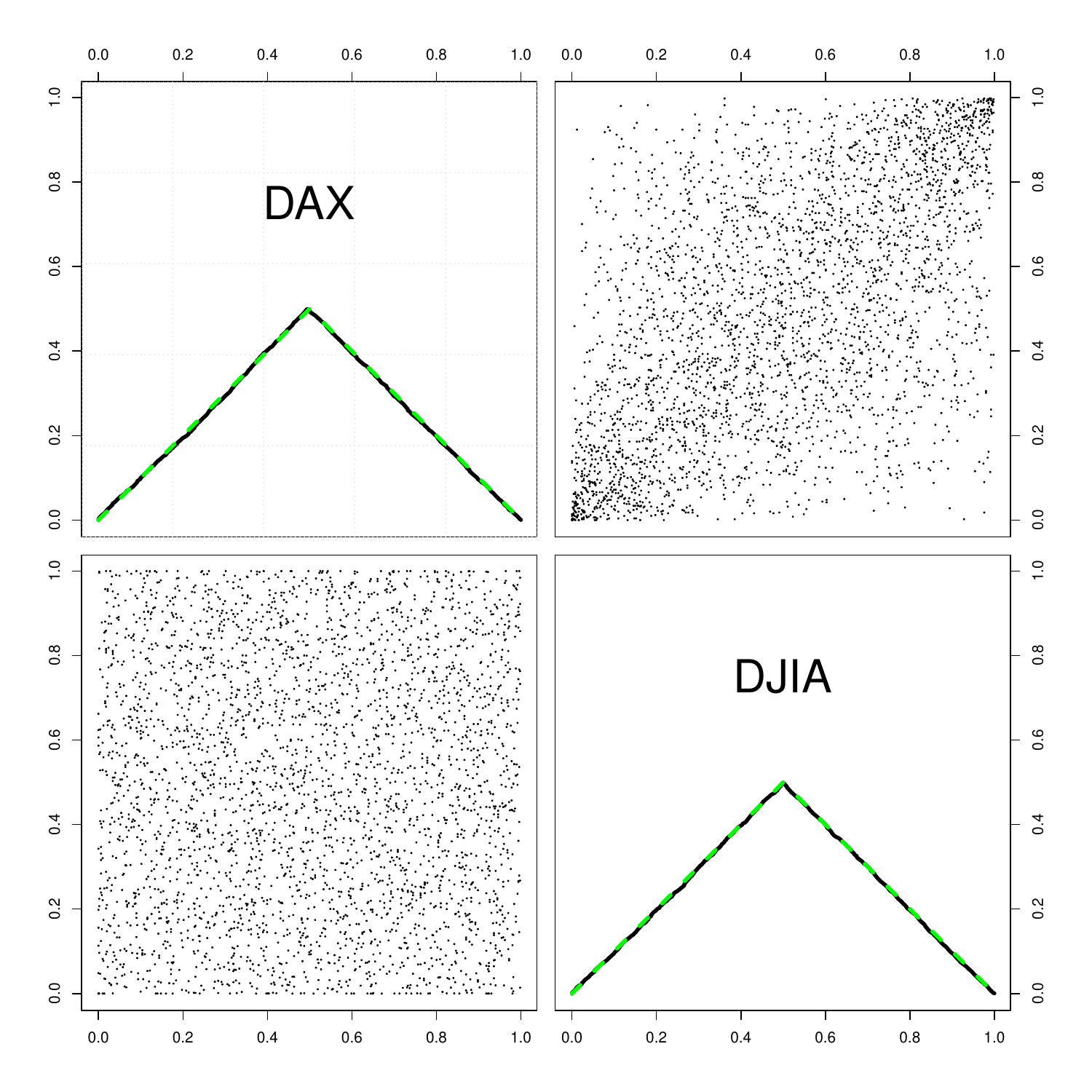} \\
	\end{minipage}%
	\hspace*{0.5cm}
	\begin{minipage}{.465\textwidth}
		\centering
		\includegraphics[width=1\linewidth]{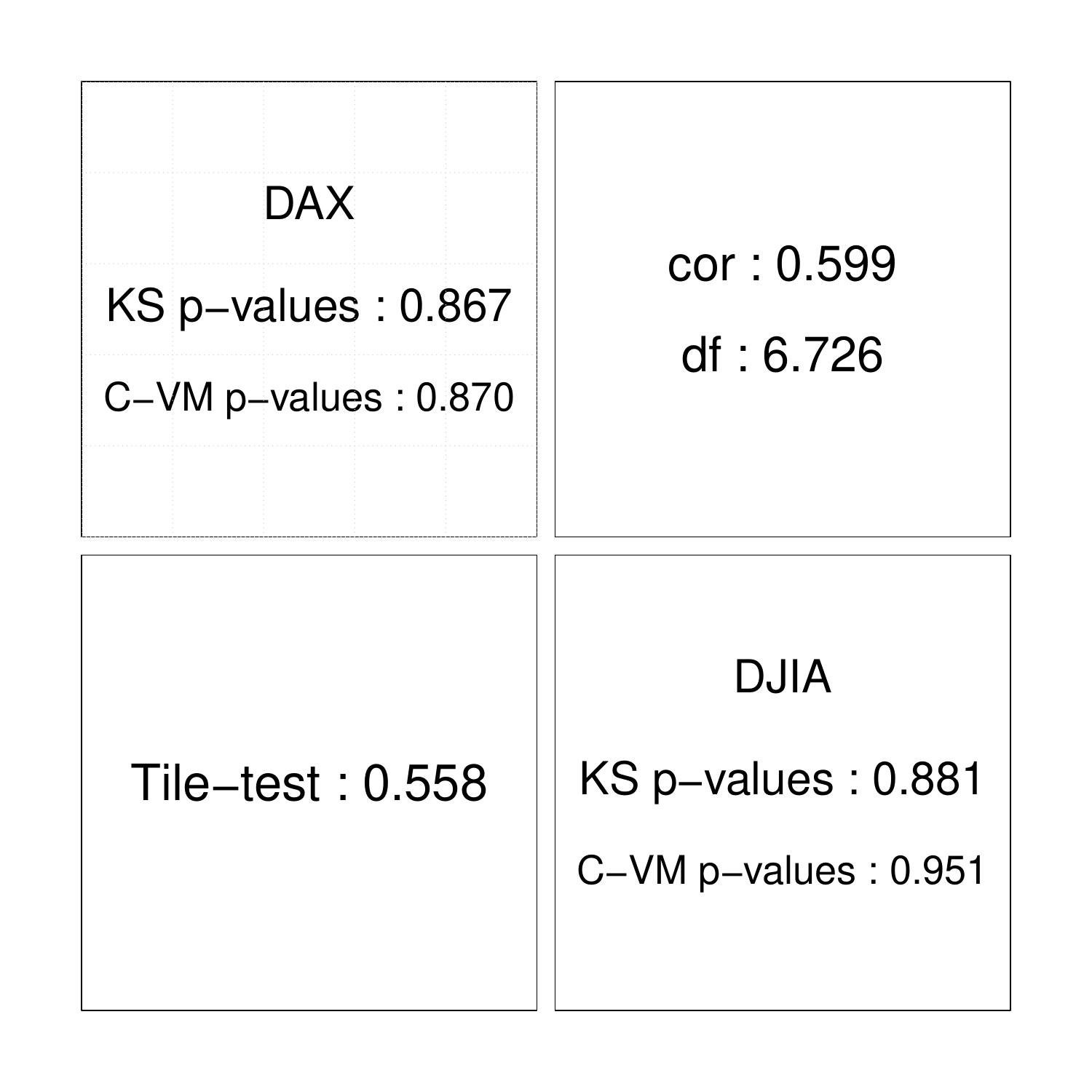}\\
	\end{minipage}
	\caption{The out-of-sample tests for the DAX and DJIA indexes.
		The diagonal on the left shows the folded distribution for the empirical probtile $z$ for each index (black) and the folded cdf for a uniform distribution (dashed green).
		The diagonal on the right shows the Kolmogorov-Smirnov and Cram\'er-von Mises $p$-values for uniformity of the marginal distribution of $z$.
		The upper-right panels show the out-of-sample empirical copula and the corresponding parameters for an estimated Student copula.
		The bottom-left panels show the out-of-sample copula after Rosenblatt transform (left), with the corresponding $p$-value for a uniformity test using the tile-test.
}
	\label{fig:probtilesOUT}
\end{figure}
Figure \ref{fig:probtilesOUT} shows the \textit{probtiles-out} set-up for DAX and DJIA, with the distributions and $p$-values for the various tests.
In this case, the tile-test gives a $p$-value of 56\%, and the uniformity cannot be rejected.
This shows that the risk evaluation using the innovations delivers an effective forecast in this bivariate case.
On the diagonal, the tests for uniformity is also largely accepted.
Notice that the $p$-values on the diagonal are too large due to the algorithm using a trailing sample, see \cite{Zumbach.TileTest} for a detailed discussion.

\FloatBarrier

\section{Out-of-sample empirical results}
\label{sec:out-of-sampleResults}
The out-of-sample framework presented in the previous section is used to study bivariate forecasts, using the same data set as in Section~\ref{sec:inSampleResults}.
Because out-of-sample, an approximate univariate uniform distribution for the probtile $z$ must result from the forecast.
With the risk algorithm described above, the Kolmogorov-Smirnov and Cram\'{e}r-von Mises tests on the marginals are always accepted by a wide margin (see however \cite{Zumbach.TileTest} for the subtle bias due to the risk algorithm).

\begin{figure}[ht]
	\centering
	\includegraphics[width=1\linewidth]{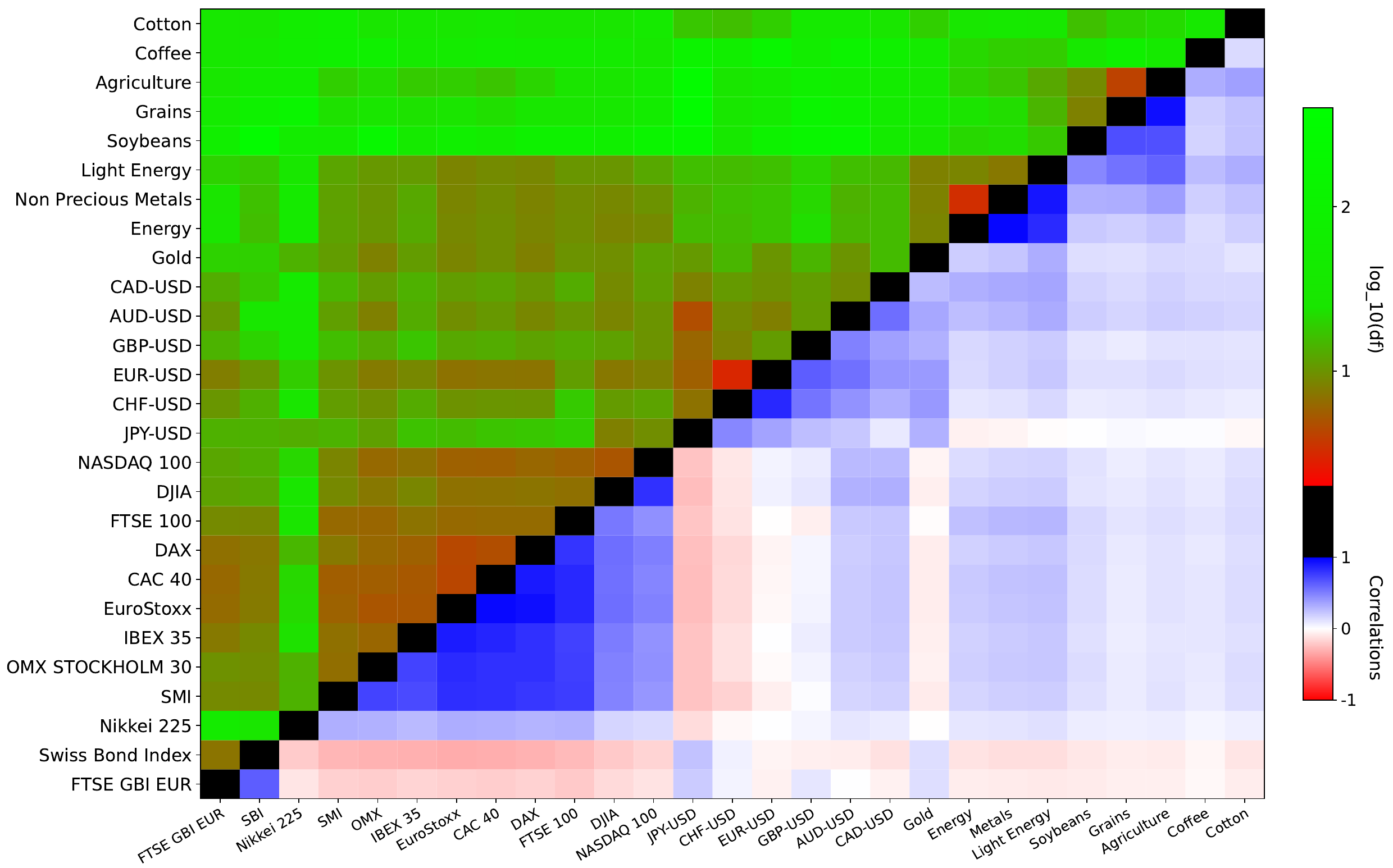}
	\caption{
		Correlations and log-degrees of freedom of an estimated bivariate out-of-sample Student copula for each pair of asset.
		The set-up is identical to Fig.~\ref{fig:df_corr_in}. }
	\label{fig:df_corr_out}
\end{figure}
First, a Student copula can be estimated on the out-of-sample probtiles.
The inverse relationship between the estimated correlations and degrees of freedom is similarly observed, as in the in-sample case in Fig.~\ref{fig:df_corr_in}.
This figure shows that very similar dependencies and estimated parameters occur in- and out-of-sample.

\begin{figure}[ht]
	\centering
	\includegraphics[width=1\linewidth]{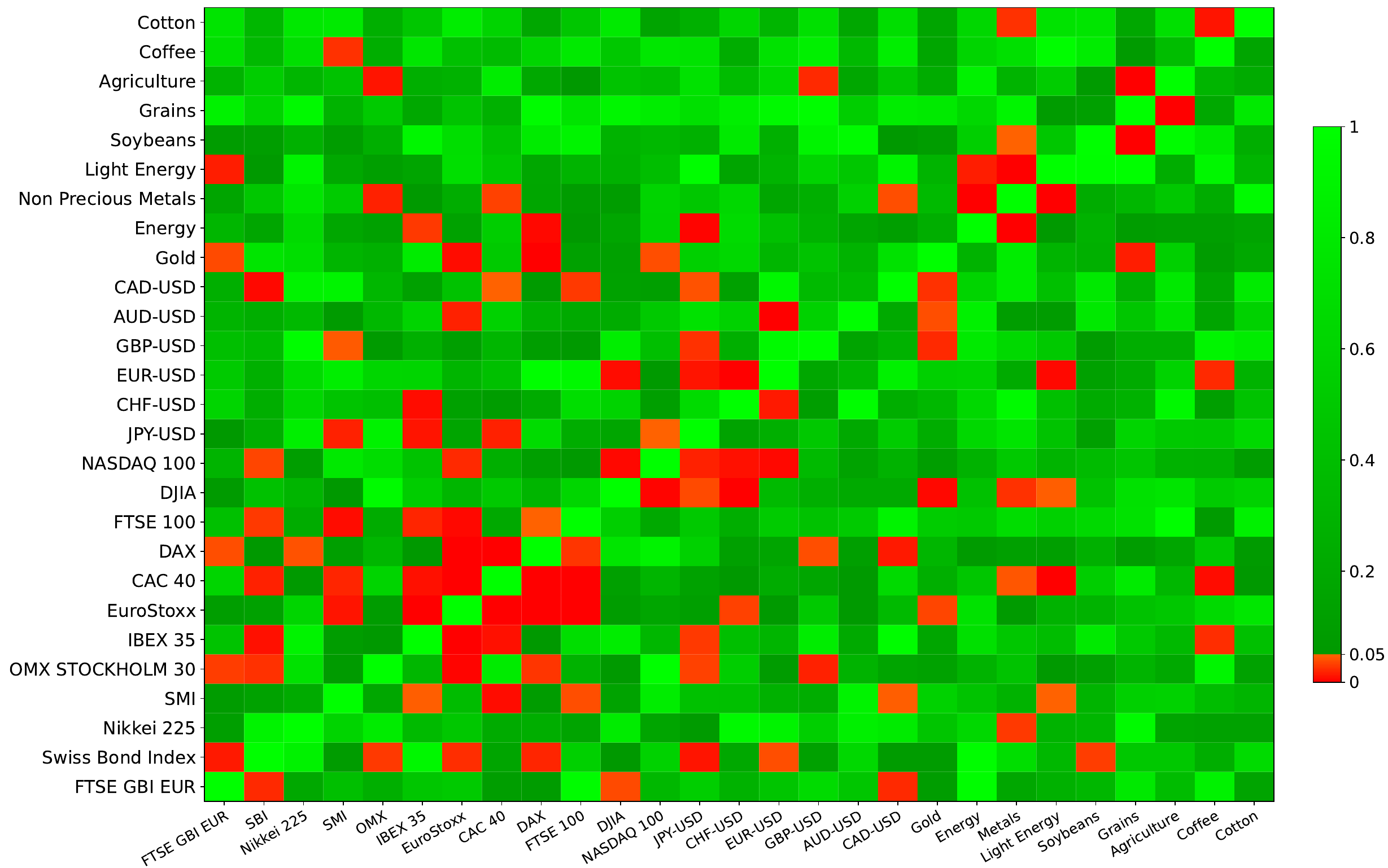}
	\caption{
		$p$-values of the tile-test out-of-sample.
		The layout is identical to Figure \ref{fig:inTile}.
	}
	\label{fig:outTile}
\end{figure}
Second, the bivariate distribution of the out-trf-copula can be tested for uniformity.
The sampled version of the Rosenblatt transform \eqref{eq:empiricalrosenblatt} is used, with the mapping \eqref{eq:inToOutCdf} to obtain the $p$-values, and where $\cdf_\text{in-sample}$ has already been computed with Monte Carlo simulations in Sec.~\ref{subsec:p-Values}.
The dependency on the sample size $n$ is treated similarly with a linear interpolation.
The results are presented with the same layout as Fig.~\ref{fig:inTile}.

The rate of rejection at the 5\% confidence level is 87/702 = 12\%.
This is too high, although still acceptable, but an out-of-sample setting is obviously more demanding.
The figure shows a similar cluster of rejections for stock indexes with high correlations, together with apparently random rejections spread over the figure.
For the rejection cases, a visual inspection of the transformed copulas shows no obvious deviations from uniformity.
Notice the sample lengths of a couple of decades, making the statistical test very discriminant.

On the optimistic side, 88\% of the dependencies are not rejected according to a rigorous statistical test, namely a historical bivariate sample of 500 innovations provides for a good forecast for the distribution of the realized innovations.
Although there is space for improvements, this overall broad validation shows that the present bivariate risk forecast captures mainly correctly the mutual dependencies between risk factors.
This result is important for the risk computations of portfolios, and for the disaggregation of a portfolio risk into its components.

\FloatBarrier
\section{Conclusions}

Risk is a forecast, and it is very important to validate a risk algorithm using backtest.
An extensive backtest for the univariate case was presented in \citep{Zumbach.TileTest} using a similar tile-test.
This paper focuses on the bivariate dependencies between major assets, both in-sample and out-of-sample in a risk forecast context.
The time-varying volatility is first discounted using a LM-ARCH process, in order to obtain the homoskedastic time series of innovations.
The validation method is first presented in the in-sample setting, where the marginals and copulas are evaluated using the full data set.
This first study is close to a textbook application of copulas, but with a Rosenblatt transform to assess the validity of the estimated copulas using $p$-values according to two statistical tests.
The key result is that Student copulas give a good description of the in-sample empirical dependencies.

In this paper, we extend this scheme to a risk forecast setting, where the information up to $t$ is used to build a forecast for the risk horizon $\DT$ and with one market outcome known at time $t+\DT$.
The univariate forecast for the cdf is used to compute the probability of a (out-of-sample) quantile using a PIT, that we called a probtile.
The bivariate forecast for the copula is used to compute a Rosenblatt transform of the probtiles to a new copula, called out-trf-copula, that should be uniform if the forecast is correct.
The $p$-value of the tile-test for uniformity can be computed, after correcting for the different set-up by using Monte Carlo simulations.
These test values show that the bivariate forecasts are good, and in particular the dependencies are captured correctly to a large extend by an empirical pdf based on a moving sample of innovations.

These out-trf-copulas and the associated statistical tests confirm that risk evaluations based on a trailing sample of historical innovations provide for good bivariate forecasts, regardless of the major type of assets and of the correlation values.
This is important for risk evaluation in finance, both when aggregating individual positions in a portfolio, and when computing finer risk diagnostics in order to mitigate the risk exposure.
Yet, this forecast algorithm is not perfect, in particular between strongly correlated time series.
The methodology presented in this paper can be used in order to backtest and improve risk algorithms.

Two statistical tests have been used to check for uniformity of the in-trf-copulas and out-trf-copulas, using Monte Carlo simulations to calibrate the tests.
In the out-of-sample setting, the test of \citep{GenestRemillard.2004} has a dependency on the correlation of the original copula, making it unsuitable for applications.
Instead, the tile-test is robust against changing the correlation, at least to a large extend.
Possibly, more studies could be done in order to assess the statistical properties of uniformity tests in a setting similar to our empirical out-of-sample computations.

The Rosenblatt transform involves a conditional expectations using the forecasted copula, which is known on a sample.
Because of the historical sample size used to build the risk forecast (i.e. 500 points), only the bivariate case can be studied.
Similarly, because historical time series are available for a couple of decades, only a risk horizon of 1 day can be analyzed in the bivariate setting.
These limitations in the multivariate direction and in the risk horizon direction are not fundamental, and the tests presented here could in principle be applied to larger spaces and longer risk horizons.
Yet, the practical computations impose the above strong limits.
Moving beyond those limits would require at least a factor 10 in the sample lengths, which is unrealistic.
Yet, despite this limited testing range, the risk methodology used in this paper essentially passes the statistical tests.
A rejection would have cast a strong shadow about the validity of the present underlying risk evaluation.

Some industries like pension funds or insurances impose some quite different parameters for risk evaluations, like a 99\% VaR at a risk horizon of 1 year.
From the industry point-of-view, these risk parameters make sense, but from a statistical rigorous validation setting, such parameters are unrealistic.
A rule of thumb from statisticians is that at least 30 points are needed to compute a statistic.
With this order of magnitude, assuming that 30 exceedances are required for statistical testing, a sample of 3000 years is needed.
Clearly, there is no way that a validation could be done for these VaR parameters, and one must only ``trust'' the results of such long term risk evaluation.
With the amount of data presently available, we have to limit ourself to the more modest risk specifications used in this document in order to make a fundamental evaluation of a risk methodology.
Yet, a failure already at this level would raise doubts about the capability of a risk methodology to operate correctly with more extreme parameters.

Finally, a cross sectional study of the innovations for financial time series is done using Student copulas, for all the pairs in a sample of 27 major indexes and FX, both in-sample and out-of-sample.
The results are similar in both settings, with the estimated parameters showing an inverse relationship between the correlation and the number of degrees of freedom.
Essentially, asset pairs with larger correlations tend to have more extreme co-movements.
This dependency, together with the changing correlations, point to a quite complex multivariate copula in finance.
As a first approximation, a multivariate Student with $df = 6$ should provide for a simple and easily estimated distribution, and at least better than a multivariate normal distribution.


\bibliographystyle{plainnat}
\bibliography{bibliography, ../../../Shared/LaTeX/biblioUniverse}

\end{document}